\documentclass[12pt,a4]{article}

\usepackage{amsmath,amssymb,amsfonts,amsthm,hyperref,bbm,latexsym,epsfig}
\usepackage{graphicx,multirow}
\usepackage{color}
\usepackage[utf8]{inputenc}
\usepackage{cite}

\definecolor{violet}{rgb}{0.54,0.17,0.9}

\newcommand{\bs}{\begin{subequations}}
\newcommand{\es}{\end{subequations}}
\newcommand{\be}{\begin{equation}}
\newcommand{\ee}{\end{equation}}
\newcommand{\ba}{\begin{eqnarray}}
\newcommand{\ea}{\end{eqnarray}}
\newcommand{\no}{\nonumber \\}

\newcommand{\tb}{\tan{\beta}}
\allowdisplaybreaks
\begin{document}

\title{
{ \normalsize\hfill}
  \\
  \LARGE One-loop corrections to the $Zb\bar{b}$ vertex in models
  with scalar doublets and singlets}

\author{
Duarte Fontes,$^{(1)}$\thanks{\tt duarte.fontes@tecnico.ulisboa.pt}
\
Lu\'is~Lavoura,$^{(1,2)}$\thanks{\tt balio@cftp.tecnico.ulisboa.pt}
\
Jorge C.\ Rom\~ao,$^{(1,2)}$\thanks{\tt jorge.romao@tecnico.ulisboa.pt}
\\ and\
Jo\~ao P.\ Silva$^{(1,2)}$\thanks{\tt jpsilva@cftp.tecnico.ulisboa.pt}
\\*[3mm]
$^{(1)} \! $
\small Universidade de Lisboa, Instituto Superior T\'ecnico, CFTP, \\
\small Av.~Rovisco Pais~1, 1049-001 Lisboa, Portugal
\\[2mm]
$^{(2)} \! $
\small Universidade de Lisboa, Instituto Superior T\'ecnico,
Departamento de F\'\i sica, \\
\small Av.~Rovisco Pais~1, 1049-001 Lisboa, Portugal
\\*[2mm]
}

\date{\today}

\maketitle
\begin{abstract}
  We study the one-loop corrections to the $Zb\bar{b}$ vertex in
  extensions of the Standard Model with arbitrary numbers of scalar
  doublets, neutral scalar singlets, and charged scalar singlets.
  Starting with a general parameterization of theories with neutral
  and singly-charged scalar particles,
  we derive the
  {conditions that, in a renormalizable model,}
  must be obeyed by the couplings in order for the divergent
  contributions to cancel.  Then, we show that those
  {conditions are indeed}
  obeyed by the models that we are interested in, and we write down
  the full finite expression for the vertex in those models.  We apply
  our results to some particular cases,
  highlighting the importance of {the}
  diagrams with neutral
  scalars{.}
\end{abstract}

\newpage

\maketitle

\section{\label{sec:intro}Introduction}

The discovery of a scalar particle at the
LHC~\cite{Aad:2012tfa,Chatrchyan:2012ufa} urges the questions of
whether there are more neutral 
scalars and whether there are charged scalars.  Multi-scalar models
have long been studied---for reviews see, for example,
Refs.~\cite{gunion:1989we,Branco:2011iw, Ivanov:2017dad}.  Here, we concentrate on
models with $n_d$ scalar doublets, $n_c$ charged-scalar singlets,
and $n_n$ neutral-scalar singlets.
The scalar-particle content is, thus,
$2n \equiv 2 \left( n_d + n_c \right)$ charged scalars $H_a^\pm$
$(a = 1, \dots, n)$ and $m \equiv 2 n_d + n_n$ neutral scalars
$S_l^0$ $(l = 1, \dots, m)$.
(The $S_l^0$ are real fields.) In our notation,
$H_1^\pm = G^\pm$ and $S_1^0 = G^0$
are, respectively, the charged and neutral would-be Goldstone bosons.

Light extra scalars may be detected directly through their production,
while heavy scalars may be detected indirectly
through their impact on the radiative corrections. 
We focus on the coupling
$Z b \bar{b}$:\footnote{We use the conventions of Ref.~\cite{Romao:2012pq},
  taking all the $\eta$ signs to be positive.
  In our convention, $g = e \left/ s_W \right.$.}
\be
\mathcal{L}_{Zbb} = -
\frac{g}{c_W}\, Z_\lambda\, \bar b\, \gamma^\lambda
\left( g_{Lb} P_L + g_{Rb} P_R \right) b,
\label{Zbb}
\ee
where $P_{L,R}$ are the projectors of chirality and, at the tree
level,
\be
g_{Lb}^0 = \frac{s_W^2}{3} - \frac{1}{2},
\quad \quad
g_{Rb}^0 = \frac{s_W^2}{3}
\label{gLb_gRb}
\ee
in models without extra gauge fields.  As usual, $s_W$ and $c_W$ are
the sine and the cosine, respectively, of the Weinberg angle
$\theta_W$.

Haber and Logan \cite{Haber:1999zh} have considered the one-loop
corrections to the vertex $Zb \bar{b}$ in models with extra scalars in
any representation of the gauge group $SU(2)_L$. The one-loop
corrected couplings can conveniently be written as
\begin{equation}
  \label{eq:5}
  g_{\aleph b} = g_{\aleph b}^\textrm{SM} + \delta g_{\aleph b}
\quad \quad (\aleph=L,R),
\end{equation}
where $g_{\aleph b}^\textrm{SM}$ includes the SM contributions and the
quantities $\delta g_{\aleph b}$ contain the New Physics
contributions.  Experimentally these couplings are obtained from the
measurable quantities
\begin{equation}
  \label{eq:6}
  R_b =
\frac{\Gamma \left( Z \rightarrow b \bar{b} \right)}{\Gamma
  \left( Z \rightarrow \textrm{hadrons} \right)}, \quad \quad
A_b=\frac{4}{3}\, A_{LR}^{FB} \left( b \right),
\end{equation}
where $A_{LR}^{FB}$ is the forward--backward asymmetry
measured in the process $e^- e^+ \to b \overline{b}$.
The present values for
these quantities are within 1$\sigma$ of the SM
predictions~\cite{Tanabashi:2018oca}; therefore,
studying the one-loop corrections to the $Zb\overline{b}$ vertex
can be used to constrain New Physics.
The work of Ref.~\cite{Haber:1999zh} has been used to constrain
various two-Higgs-doublet models (2HDM)\cite{Dorsch:2013wja, Basler:2016obg,
Krause:2016xku, Fontes:2015gxa,Mader:2012pm, Belusca-Maito:2016dqe},
the
Georgi--Machacek
model\cite{Campbell:2016zbp, Hartling:2014aga,
Hartling:2014zca, Chiang:2014bia, Degrande:2015xnm}, scotogenic
models\cite{Tang:2017rhv}, models with $SU(2)_L$ singlet
scalars~\cite{vonBuddenbrock:2018xar, Han:2017etg}, and used in fitting
programs\cite{Flacher:2008zq, Haller:2018nnx}.

In this paper, we extend the analysis of Ref.~\cite{Haber:1999zh}
by considering CP-violating
scalar sectors and we write down
the final results in models with singlets and doublets
in a simple and usable form.  This is possible due to a convenient
parameterization that was introduced in Refs.~\cite{Grimus:2002ux,
Grimus:2007if, Grimus:2008nb}, following earlier
work~\cite{Grimus:1989pu}. We also discuss in detail the
renormalization of the vertex for these generic models, which was
assumed but not explicitly displayed in Ref.~\cite{Haber:1999zh}.

We present the Lagrangian and the relevant calculations {in
Section~\ref{sec:calc}.}  In Section~\ref{sec:doubsing} we introduce
the parameterization relevant for doublets and singlets; we show that
all the divergences cancel out and we simplify the final expressions.
The connection with experiment is reviewed in Section~\ref{sec:exp},
and then applied in Section~\ref{sec:particular} to some simple cases,
looking in particular at the importance of diagrams with neutral
scalars.  We draw our conclusions in Section~\ref{sec:conclusions}.
An appendix summarizes the definitions of the Passarino--Veltman
functions used in this paper.

\section{\label{sec:calc}The one-loop calculation}

We use the approximation where the CKM matrix element $V_{tb} = 1$,
requiring us to consider only the quarks bottom with mass $m_b$ and
top with mass $m_t$.  We neglect $m_b$ in the propagators and loop
functions, but we keep generic couplings.

\subsection{\label{subsec:coup}Couplings}
 
In addition to the couplings in Eqs.~\eqref{Zbb} and~\eqref{gLb_gRb},
we need
\ba
\mathcal{L}_{Ztt} &=& -
\frac{g}{c_W}\, \bar t\, \gamma^\lambda
\left( g_{Lt} P_L + g_{Rt} P_R \right) t\, Z_\lambda,
\label{Ztt}
\\*[1mm]
\mathcal{L}_{Wtb} &=&  -
\frac{g}{\sqrt{2}}\, 
\left( \bar{t} \gamma^\lambda P_L b\, W_\lambda^+
+ \bar{b} \gamma^\lambda P_L t\, W_\lambda^-
\right).
\label{Wtb}
\ea
In Eq.~\eqref{Ztt},
at the tree level
\be
g_{Lt}^0 = \frac{1}{2} - \frac{2 s_W^2}{3},
\quad \quad
g_{Rt}^0 = - \frac{2 s_W^2}{3}.
\label{gLt_gRt}
\ee
From Eqs.~\eqref{gLb_gRb} and~\eqref{gLt_gRt},
\be
g_{Rb}^0 - g_{Lt}^0 = g_{Lb}^0 - g_{Rt}^0 = \frac{s_W^2 - c_W^2}{2}.
\label{difference}
\ee
The charged scalars $H_a^\pm$ and the neutral scalars $S_l^0$
interact with the quarks through
\ba
\mathcal{L}_{Htb} &=&
\sum_{a=1}^n \left[
  H_a^+\, \bar t \left( c_a^\ast P_L - d_a P_R \right) b + H_a^-\, \bar b \left( c_a P_R - d_a^\ast P_L \right) t
\right],
\label{Htb}
\\ 
\mathcal{L}_{Sbb} &=& \sum_{l=1}^m S_l^0\,
\bar b \left( r_l P_R + r_l^\ast P_L \right) b,
\label{Sbb}
\ea
and with the $Z$ gauge boson through
\ba
\mathcal{L}_{ZHH} &=&  -
\frac{g}{c_W}\, Z_\lambda
\sum_{a, a^\prime = 1}^n X_{a a^\prime}
\left( H_a^+\, i \partial^\lambda H_{a^\prime}^-
- H_{a^\prime}^-\, i \partial^\lambda H_a^+ \right),
\label{ZHH}
\\ 
\mathcal{L}_{ZSS} &=&
\frac{ i g}{c_W}\, Z_\lambda
\sum_{l, l^\prime = 1}^m Y_{l l^\prime}
\left( S_l^0\, i \partial^\lambda S_{l^\prime}^0
- S_{l^\prime}^0\, i \partial^\lambda S_l^0 \right),
\label{ZSS}
\\
\mathcal{L}_{ZZS} &=&
\frac{g M_Z}{2 c_W}\, Z_\lambda Z^\lambda \sum_{l= 1}^m y_l S_l^0,
\label{ZZS}
\ea
where $M_Z$ is the mass of the $Z$.
In general,
the coefficients $c_a$,
$d_a$,
and $r_l$ in Eqs.~\eqref{Htb} and~\eqref{Sbb} are complex,
while the $y_l$ in Eq.~\eqref{ZZS} are real.
The $n \times n$ matrix $X$ in Eq.~\eqref{ZHH} is Hermitian.
The $m \times m$ matrix $Y$ in Eq.~\eqref{ZSS} is real and antisymmetric.
We let $m_a$ denote the mass of $H_a^\pm$ and $m_l$ denote the mass of $S^0_l$.

\subsection{\label{subsec:1loop}One-loop diagrams}

\begin{figure}[htb]
  \centering
  \begin{tabular}{cc}
  \includegraphics[scale=0.8]{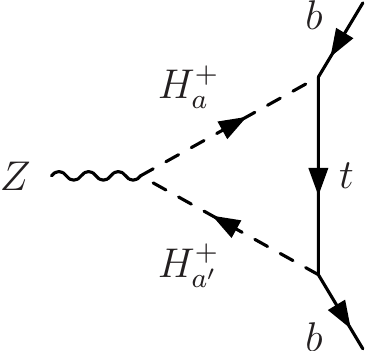}&\hskip 5mm
  \includegraphics[scale=0.8]{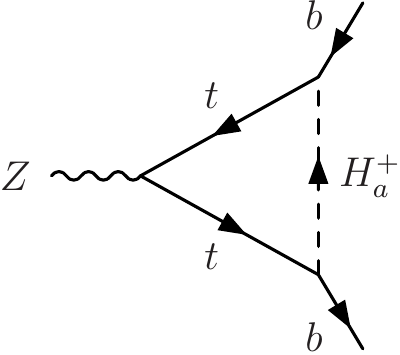}\\
  a) & \hskip 5mm b)   
  \end{tabular}
    \caption{Two diagrams with charged scalars contributing to
      the $Zb\bar{b}$ vertex.}
  \label{fig:Generic-Charged}
\end{figure}

\begin{figure}[htb]
  \centering
  \begin{tabular}{cc}
  \includegraphics[scale=0.8]{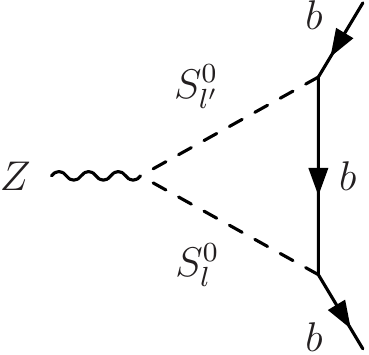}&\hskip 5mm
  \includegraphics[scale=0.8]{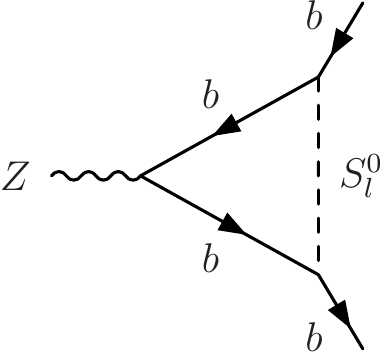}\\
  a) & \hskip 5mm b)
  \end{tabular}
    \caption{Two diagrams with neutral scalars contributing to
      the $Zb\bar{b}$ vertex.}
  \label{fig:Generic-Neutral}
\end{figure}

\begin{figure}[htb]
  \centering
  \begin{tabular}{cccc}
  \includegraphics[scale=0.8]{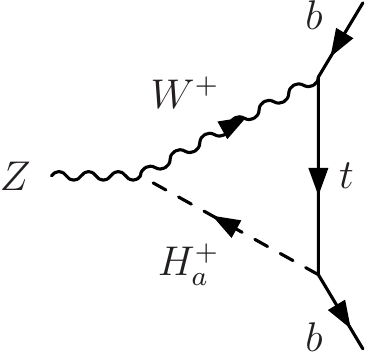}&\hskip 5mm
  \includegraphics[scale=0.8]{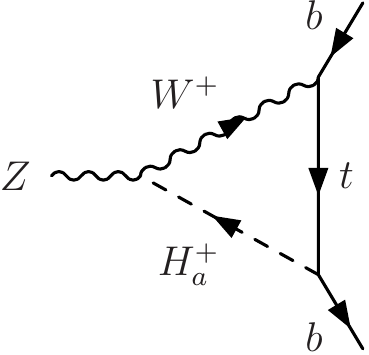}& \\
  $a$) & \hskip 5mm $b$) \\[5mm] 
  \includegraphics[scale=0.8]{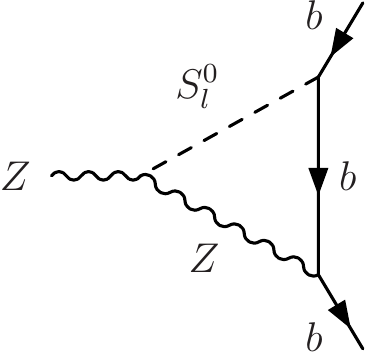}&\hskip 5mm
  \includegraphics[scale=0.8]{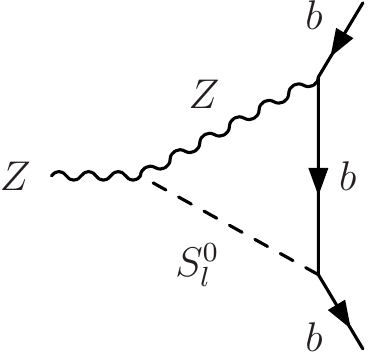} \\
  $c$) & \hskip 5mm $d$)
  \end{tabular}
  \caption{Diagrams referred as to ``type d)'' in  Ref.~\cite{Haber:1999zh}.} 
  \label{fig:type_d)}
\end{figure}

\begin{figure}[htb]
  \centering
   \begin{tabular}{ccc} 
    \includegraphics[width=0.25\textwidth]{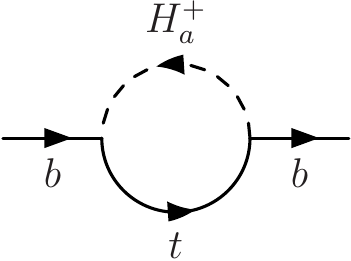}&\hskip 35mm&
    \includegraphics[width=0.25\textwidth]{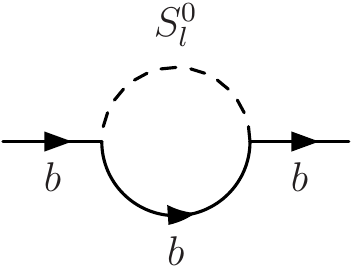}\\[+1mm]
  $a$) && $b$)
     \end{tabular}
   \caption{Contribution of the charged and neutral scalars
     to the self-energy of the bottom quark,
     leading to ``type c)'' contributions to the vertex.}
  \label{fig:Self-Energies}
\end{figure}

At one-loop level,
the diagrams contributing to the $Zb\bar{b}$ vertex
are shown in Figs.~\ref{fig:Generic-Charged} and~\ref{fig:Generic-Neutral},
for charged and neutral scalars,
respectively.
This classification of the diagrams was proposed in Ref.~\cite{Haber:1999zh},
wherein the diagrams in Fig.~\ref{fig:type_d)} were also mentioned,
but then neglected.
The diagrams in Fig.~\ref{fig:type_d)} involving the charged scalars
do not give new contributions beyond the Standard Model (SM)
in models with only scalar singlets and doublets,
because in these models there are no $Z W^\pm H_a^\mp$ couplings
other than the $Z W^\pm G^\mp$ already present in the SM.
The diagrams in Fig.~\ref{fig:type_d)}
involving neutral scalars are proportional to $m_b$.
This is because the coupling of the $Z$
to the bottom quarks in Eq.~\eqref{Zbb} conserves chirality,
\textit{i.e.}\ the ingoing and outgoing bottom quarks have the same chirality,
while the analogous coupling of a neutral scalar does not contain
the matrix $\gamma^\lambda$ and therefore
it changes the chirality of the bottom quark.
Hence,
in the diagrams in Fig.~\ref{fig:type_d)}c),d)
there must be a mass insertion
in the internal bottom-quark propagator in order to change the chirality
of the bottom-quark line once again.
Since the diagrams in Fig.~\ref{fig:type_d)} are convergent,
one may neglect them by taking $m_b=0$,
and this is what was done in Ref.~\cite{Haber:1999zh}.
Nevertheless,
because $m_b$ could appear multiplied by a large coefficient
(such as $\tan{\beta}=v_2/v_1$ in the $\mathbbm{Z}_2$-symmetric 2HDM,
see for instance Table 2 in Ref.~\cite{Branco:2011iw})
we will also present their calculation in order to
  check the validity of this approximation.
The diagrams in Figs.~\ref{fig:Generic-Charged} and~\ref{fig:Generic-Neutral}
are divergent and must be renormalized.
We follow the on-shell renormalization scheme
of Hollik~\cite{Hollik:1988ii, Hollik:1993cg}.
Applying multiplicative renormalization,
the renormalized vertex acquires some terms
leading to a correction to the $Z$ propagator;
these are part of the oblique parameters
and were shown to be very small in Ref.~\cite{Haber:1999zh}.
Here we are looking for the terms that change the tree-level couplings,
which after renormalization may be written as
\begin{equation}
  \label{eq:57}
  i \hat  \Gamma_\mu^{Zff} \! = \!
-i \gamma_\mu \frac{g}{c_W}
  \!\left[ \left( g_{Lb}^{0} + \Delta g_L \right) \! P_L \!
    + \left( g_{Rb}^{0} + \Delta g_R \right) \! P_R \right],
\end{equation}
where $\Delta g_\aleph$
($\aleph = L, R$)
represent all the one-loop corrections after renormalization,
including the ones involving $G^\pm$,
$G^0$,
and the already-observed neutral scalar with mass 125\,GeV
(more on this in Section~\ref{sec:exp}).
To perform the renormalization
one needs to evaluate the renormalization constants
that are obtained from the self-energies.
We therefore need to evaluate the contributions of
both the charged and neutral scalars to the self-energies,
shown in Fig.~\ref{fig:Self-Energies}.
The self-energy $i \Sigma \left( p \right)$
receives contributions proportional to $\not \! p P_L$,
$\not \! p P_R$,
$m_b P_L$,
and $m_b P_R$.
In our approximation of neglecting $m_b$,
we write
\be
\Sigma (p) = \not \! p \left[
  \Omega_L \left( p^2 \right) P_L + \Omega_R \left( p^2 \right) P_R \right].
\ee
Following Hollik's renormalization scheme \cite{Hollik:1988ii, Hollik:1993cg},
the self-energy produces contributions to $\Delta g_{Lb}$ and $\Delta g_{Rb}$
given by
\bs
\ba
\Delta g_{Lb} \left( c \right)
&=& - g_{Lb}^0\, \Omega_L \left( p^2 = m_b^2 \right),
\\
\Delta g_{Rb} \left( c \right) &=&
- g_{Rb}^0\, \Omega_R \left( p^2 = m_b^2 \right).
\ea
\es
Note that Ref.~\cite{Haber:1999zh} follows an equivalent procedure,
ignoring renormalization
and calculating simply the reducible diagrams
with self-energy corrections in the external bottom quarks,
which they dub ``type c) diagrams''.
Although we do perform the renormalization,
we will name the contributions arising from it as ``type c)'',
allowing for an easy comparison with Ref.~\cite{Haber:1999zh}.

Our calculations of the various diagrams
have been performed
by hand and then confirmed through the standard computer codes
\texttt{FeynRules}~\cite{Christensen:2008py},
\texttt{QGRAF}~\cite{Nogueira:1991ex},
and \texttt{FeynCalc}~\cite{Mertig:1990an, Shtabovenko:2016sxi}.
Recently,
two of us (DF and JCR) have developed the new software
\texttt{FeynMaster}~\cite{Fontes:2019wqh} that handles,
in an automated way,
all these steps.
The results involve Passarino--Veltman loop functions~\cite{Passarino:1978jh};
our conventions for them coincide with those in \texttt{FeynCalc}
and \texttt{LoopTools}~\cite{Hahn:1998yk, vanOldenborgh:1990yc},
and are summarized in Appendix~\ref{app:PV}.

We next turn to the computation of each diagram.

\subsection{\label{subsec:calc-Hpm}Calculating the diagrams
  involving charged scalars}

The diagrams in Fig.~\ref{fig:Generic-Charged}a) lead to
\bs
\label{jdoipsa}
\ba
\Delta g_{Lb} \left( a \right) &=&
\frac{1}{8\pi^2}
\sum_{a,a'=1}^{n} c_a X_{aa'} c_{a'}^\ast\,
C_{00} \left( M_Z^2, 0, 0, m_{a'}^2, m_a^2, m_t^2 \right),
\label{eq:15333}
\\
\Delta g_{Rb} \left( a \right) &=&
\Delta g_{Lb} \left( a \right) \left( c_a \rightarrow d_a^\ast \right),
\label{eq:154}
\ea
\es
where $C_{00}$ is a Passarino--Veltman function
defined through Eq.~\eqref{c00}.
We have set $m_b=0$ inside all the Passarino--Veltman functions;
however,
when evaluating them numerically it is sometimes better to keep $m_b \neq 0$
in order to avoid numerical instabilities.
We should note that the sums
in Eqs.~\eqref{jdoipsa}
start at $a=1$,
\textit{i.e.}\ they include the charged Golsdtone bosons $G^\pm$.
However,
one may show that $X_{1a}=X_{a1}=0$,
and therefore the sum in Eq.~(\ref{eq:15333})
may start at $a, a^\prime = 2$,
while the term with $a = a^\prime = 1$
is separately included in the SM contribution.

The diagrams in Fig.~\ref{fig:Generic-Charged}b) lead,
after
taking into account that
\begin{equation*}
\left( d - 2 \right) C_{00} \left( \dots \right)
= 2\, C_{00} \left( \dots \right) - 1/2
\end{equation*}
($d$ is the dimension of space--time),
to
\bs
\ba
\Delta g_{Lb} \left( b \right) &=&
\frac{1}{16 \pi^2} \sum_{a=1}^{n} \left| c_a \right|^2 \left\{
{\vbox to 14 pt{}}
- m_t^2\, g_{Lt}^0\, C_0 \left( 0, M_Z^2, 0, m_a^2, m_t^2, m_t^2 \right)
\right.
\nonumber \\ & &
+ g_{Rt}^0 \bigg[
{\vbox to 12 pt{}}
2\, C_{00} \left( 0, M_Z^2, 0, m_a^2, m_t^2, m_t^2 \right)
- \frac{1}{2}
\nonumber \\ & & \left.
- M_Z^2\, C_{12} \left( 0, M_Z^2, 0, m_a^2, m_t^2, m_t^2 \right)
{\vbox to 12 pt{}}
\bigg]
{\vbox to 14 pt{}}
\right\} \! ,
\label{eq:164a}
\\*[1mm] 
\Delta g_{Rb} \left( b \right) &=& \Delta g_{Lb} \left( b \right)
\left( c_a \rightarrow d_a,\ g_{Lt}^0 \leftrightarrow g_{Rt}^0 \right).
\label{eq:164b}
\ea
\es
The Passarino--Veltman function $C_0$ is defined in Eq.~\eqref{c0},
while $C_{12}$ is defined through Eq.~\eqref{c00}.

As for the type c) contributions,
arising through renormalization from
the diagram in Fig.~\ref{fig:Self-Energies}a),
we find
\bs
\label{eq:158}
\ba
\Delta g_{Lb} \left( c \right) &=&
\frac{g_{Lb}^0}{16 \pi ^2}
\sum_{a=1}^{n} \left| c_a \right|^2 
B_1 \left( 0, m_t^2, m_a^2 \right),
\label{eq:157}
\\
\Delta g_{Rb} \left( c \right) &=&
\Delta g_{Lb} \left( c \right) \left( c_a \rightarrow d_a,\
g_{Lb}^0 \to g_{Rb}^0 \right).
\label{eq:15777}
\ea
\es
The Passarino--Veltman function $B_1$ is defined in Eq.~\eqref{b1}.

In the CP-conserving limit,
Eqs.~\eqref{jdoipsa}--\eqref{eq:158} agree with Eqs.~(4.1)
of Ref.~\cite{Haber:1999zh},
and also with Ref.~\cite{Kundu:1995qb}.

The functions $B_1$ and $C_{00}$ are divergent;
all the other Passarino--Veltman functions appearing in this paper are finite.
In dimensional regularization,
defining the divergent quantity
\be
\textrm{div} = \frac{2}{4 - d} - \gamma + \ln{\left( 4 \pi \right)},
\label{divdef}
\ee
one has
\bs
\ba
B_1 \left( r^2, m_0^2, m_1^2 \right) &=& - \frac{\textrm{div}}{2} +
\mathrm{finite\ terms},
\\
C_{00} \left[ r_1^2, \left( r_1 - r_2 \right)^2, r_2^2,
m_0^2, m_1^2, m_2^2 \right] &=&
+ \frac{\textrm{div}}{4} + \mathrm{finite\ terms}.
\label{eq:4}
\ea
\es
Therefore,
the divergent terms in Eqs.~\eqref{jdoipsa}--\eqref{eq:158} are
\bs
\label{eq:166}
\ba
& & \Delta g_{Lb} \left( a \right) + \Delta g_{Lb} \left( b \right)
+ \Delta g_{Lb} \left( c \right) \no
&=& \frac{\textrm{div}}{32 \pi^2}
\left[
\sum_{a,a'=1}^{n} c_a X_{aa'} c_{a'}^\ast
+ \left( g_{Rt}^0 - g_{Lb}^0 \right) \sum_{a=1}^{n} \left| c_a \right|^2
\right] + \cdots ,
\\
& & \Delta g_{Rb} \left( a \right) + \Delta g_{Rb} \left( b \right)
+ \Delta g_{Rb} \left( c \right) \no
&=& \frac{\textrm{div}}{32 \pi^2} \left[
\sum_{a,a'=1}^{n} d_a^\ast X_{aa'} d_{a'}
+ \left( g_{Lt}^0 - g_{Rb}^0 \right) \sum_{a=1}^{n} \left| d_a \right|^2
\right] + \cdots.
\ea
\es
We thus conclude that in any sensible theory one must have
\bs
\label{eq:165}
\ba
\sum_{a,a'} c_a X_{aa'} c_{a'}^\ast
&=&
\frac{s_W^2 - c_W^2}{2} \sum_{a} \left| c_a \right|^2,
\label{eq:165a}
\\
\sum_{a,a'} d_a^\ast X_{aa'} d_{a'}
&=&
\frac{s_W^2 - c_W^2}{2} \sum_{a} \left| d_a \right|^2,
\label{eq:165b}
\ea
\es
where we have used Eq.~\eqref{difference}.

\subsection{\label{subsec:calc-S0}Calculating the diagrams
  involving neutral scalars}

The diagrams in Fig.~\ref{fig:Generic-Neutral}a) lead to
\bs
\label{eq:153}
\ba 
\Delta g_{Lb} \left( a \right) &=& 
\frac{i}{4\pi^2}
\sum_{l,l'=1}^{m} r_l Y_{ll'} r_{l'}^\ast\,
C_{00} \left( 0, M_Z^2, 0, 0, m_{l'}^2, m_l^2 \right),
\label{eq:153N}
\\
\Delta g_{Rb} \left( a \right) &=&
\Delta g_{Lb} \left( a \right) \left( r_l \rightarrow r_l^\ast \right).
\label{eq:154N}
\ea
\es
The diagrams in Fig.~\ref{fig:Generic-Neutral}b) lead to
\bs
\begin{align}
\Delta g_{Lb} \left( b \right) =&\, \frac{g_{Rb}^0}{16 \pi^2}
\sum_{l=1}^{m} \left| r_l \right|^2 
\bigg[ 2\, C_{00} \left( 0, M_Z^2, 0, m_l^2, 0, 0 \right)  - \frac{1}{2} \nonumber \\
& \hspace{10mm} - M_Z^2\, C_{12} \left( 0, M_Z^2, 0, m_l^2, 0, 0 \right)
\bigg],
\label{eq:155b}
\\
\Delta g_{Rb} \left( b \right) =&\,
\Delta g_{Lb} \left( b \right) \left( g_{Rb}^0 \to g_{Lb}^0 \right).
\label{eq:156b}
\end{align}
\es
As for the type c) contributions, arising through renormalization from
Fig.~\ref{fig:Self-Energies}b), we find
\bs
\label{eq:158N}
\ba
\Delta g_{Lb} \left( c \right) &=&
\frac{g_{Lb}^0}{16 \pi^2}
\sum_{l=1}^{m} \left| r_l \right|^2  B_1 \left( 0, 0, m_l^2 \right),
\label{eq:157N}
\\
\Delta g_{Rb} \left( c \right) &=&
\Delta g_{Lb} \left( c \right) \left( g_{Lb}^0 \rightarrow g_{Rb}^0 \right).
\label{eq:159N}
\ea
\es
In the CP-conserving limit,
Eqs.~\eqref{eq:153}--\eqref{eq:158N} agree with Eqs.~(5.1)
of Ref.~\cite{Haber:1999zh}.

Collecting all the divergent terms in Eqs.~\eqref{eq:153N},
\eqref{eq:155b},
and~\eqref{eq:157N} we find
\ba
& & \Delta g_{Lb} \left( a \right)
+ \Delta g_{Lb} \left( b \right)
+ \Delta g_{Lb} \left( c \right) \no
&=&
\frac{\textrm{div}}{32 \pi^2}
\left[ 2 i \sum_{l,l'=1}^{m} r_l Y_{ll'} r_{l'}^\ast 
+ \left( g_{Rb}^0 - g_{Lb}^0 \right) \sum_{l=1}^{m} \left| r_l \right|^2
\right] + \cdots.
\label{eq:166N}
\ea
Since $g_{Rb}^0 - g_{Lb}^0 = 1/2$,
a consistent theory requires
\be
\sum_{l,l'=1}^{m} r_l Y_{ll'} r_{l'}^*
=
 \frac{i}{4}
\sum_{l} |r_l|^2.
\label{eq:consistent_N}
\ee
This {condition}
can also be obtained by collecting all the
divergent terms in Eqs.~\eqref{eq:154N}, \eqref{eq:156b},
and~\eqref{eq:159N}.

The diagrams in Fig.~\ref{fig:type_d)}c),d) involve neutral scalars.
They are not divergent and they are
{suppressed by}
$m_b$.
However, we keep them because they might be enhanced when the coupling
of neutral scalars to the bottom quark gets enhanced, as in the
type-II 2HDM.  From them we get
\bs
\ba
\Delta g_{Lb} \left( d \right) &=&
\frac{g m_b M_Z}{8 \pi^2 c_W}
\sum_{l=1}^{m} y_l\, \mathrm{Re}\,r_l
\left\{
{\vbox to 14 pt{}}
g_{Lb}^0 \Big[ C_0 \left( M_Z^2, 0, 0, M_Z^2, m_l^2, 0 \right)
\right.
\nonumber \\ & & \left.
\hskip 30mm
- C_1 \left( M_Z^2, 0, 0, M_Z^2, m_l^2, 0 \right) \Big] \nonumber \right. \\
&& \left. \hskip 30mm + g_{Rb}^0\, C_1 \left( M_Z^2, 0, 0, m_l^2, M_Z^2, 0 \right)
{\vbox to 14 pt{}} \right\}, \hspace*{5mm}
\label{eq:16cc}
\\
\Delta g_{Rb} \left( d \right) &=&
\Delta g_{Lb} \left( d \right) \left( g_{Lb}^0 \leftrightarrow
g_{Rb}^0 \right).
\label{eq:17cc}
\ea
\es
The function $C_1$ is defined through Eq.~\eqref{c1}.

At this juncture we want to make a clarification.  The one-loop
results for $\Delta g_{Lb}$ and $\Delta g_{Rb}$ have imaginary parts.
If there are no scalars with mass below $M_Z/2$, then the imaginary
parts only appear through cuts of the internal bottom-quark lines of
Fig.~\ref{fig:Generic-Neutral}b), thus affecting only the
contributions with neutral scalars.  Although those imaginary parts
may be of the same order of magnitude as the real parts, they are
unimportant because the observables will depend on, for example,
\ba
\left| g_{Lb} \right|^2 &=&
\left| g_{Lb}^0 + \Delta g_{Lb} \right|^2
\no
&=& \left| g_{Lb}^0 \right|^2
+ 2\, \textrm{Re} \left( g_{Lb}^0\, \Delta g_{Lb}^\ast \right)
+ \mathrm{O} \left( \Delta g_{Lb}^2 \right)
\no
&=& \left| g_{Lb}^0 \right|^2
+ 2\, g_{Lb}^0\, \textrm{Re}\left( \Delta g_{Lb} \right)
+ \mathrm{O} \left( \Delta g_{Lb}^2 \right),
\ea
where the last line follows from the fact that $g_{Lb}^0$ is real.  As
a result, the impact of an imaginary $\Delta g_{Lb}$ on the
observables (see the next section) effectively appears only at higher
order.

\subsection{\label{subsec:summary}Summary}

A generic theory with the couplings in Eqs.~\eqref{Zbb},
\eqref{Ztt},
\eqref{Wtb},
and~\eqref{Htb}--\eqref{ZZS} gets radiative corrections to the $Z b \bar{b}$
vertex,
obtained at the one-loop level by summing our Eqs.~\eqref{eq:15333},
\eqref{eq:164a},
\eqref{eq:157},
\eqref{eq:153N},
\eqref{eq:155b},
and~\eqref{eq:157N}---and,
if enhanced,
\eqref{eq:16cc}---for $\Delta g_{Lb}$,
and by summing our Eqs.~\eqref{eq:154},
\eqref{eq:164b},
\eqref{eq:15777},
\eqref{eq:154N},
\eqref{eq:156b},
and~\eqref{eq:159N}---and,
if enhanced,
\eqref{eq:17cc}---for $\Delta g_{Rb}$.
The theory only makes sense if its couplings are related through
Eqs.~\eqref{eq:165a},
\eqref{eq:165b},
and~\eqref{eq:consistent_N},
which are needed in order for the divergences to cancel.

\section{\label{sec:doubsing}Models with doublet and singlet scalars}

We now focus on extensions of the SM with $n_d$ scalar doublets, $n_c$
singly-charged scalar $SU(2)_L$ singlets, and $n_n$ real scalar
gauge-invariant fields.  The particle content is then $2 n \equiv 2
\left( n_d + n_c \right)$ charged scalars $H_a^\pm$ and $m \equiv 2
n_d + n_n$ neutral scalars $S_l^0$; this counting includes the
Goldstone bosons $H_1^\pm = G^\pm$ and $S_1^0 = G^0$.  Without loss of
generality, one may assume that the scalar with mass 125\,GeV found at
the LHC is $S^0_2$; generality is lost if one makes the further
assumption that the masses are ordered, since there might be massive
scalar(s) below 125\,GeV.

The scalar doublets are
\be
\Phi_k = \left( \begin{array}{c} \varphi_k^+ \\*[1mm] \varphi_k^0 \end{array}
\right),
\quad \quad
\tilde \Phi_k \equiv i \sigma_2 \Phi_k^\ast
= \left( \begin{array}{c} {\varphi_k^0}^\ast \\*[1mm] - \varphi_k^- \end{array}
\right).
\ee
The fields $\varphi_k^0$ have VEVs $v_k \left/ \sqrt{2} \right.$,
where the $v_k$ may be complex.

Obviously,
the charged and neutral $SU(2)_L$ singlets have no Yukawa couplings.
The Yukawa Lagrangian is
\be
\begin{split}
\mathcal{L}_\textrm{Yukawa} = -
\left( \begin{array}{cc} \overline{t_L} & \overline{b_L} \end{array} \right)
\sum_{k=1}^{n_d} \Bigg[
& f_k
\left( \begin{array}{c} \varphi_k^+ \\*[2mm]
\varphi_k^0 \end{array} \right) b_R
+ e_k
\left( \begin{array}{c} {\varphi_k^0}^\ast \\*[1mm]
- \varphi_k^-
\end{array} \right) t_R \Bigg] + \textrm{H.c.},
\label{jcksps}
\end{split}
\ee
where the $e_k$ and the $f_k$ ($k = 1, \ldots, n_d$)
are the Yukawa coupling constants.

\subsection{Formalism}

We use the formalism
in Refs.~\cite{Grimus:2002ux, Grimus:2007if, Grimus:2008nb}.
We write $\varphi_k^+$ and $\varphi_k^0$ as superpositions
of the physical (i.e. eigenstates of mass) fields as
\ba
\varphi_k^+ &=& \sum_{a=1}^n \mathcal{U}_{ka} H_a^+,
\label{jdkpsos} \\
\varphi_k^0 &=& \frac{1}{\sqrt{2}}
\left( v_k + \sum_{l=1}^m \mathcal{V}_{kl} S_l^0 \right).
\label{def:U&V}
\ea
The matrix $\mathcal{U}$ is $n_d \times n$
and the matrix $\mathcal{V}$ is $n_d \times m$.

Since $H_1^\pm$ and $S_1^0$ are Goldstone bosons,
the first columns of $\mathcal{U}$ and $\mathcal{V}$ are fixed and given by
\be
\mathcal{U}_{k1} = \frac{v_k}{v},
\quad \quad
\mathcal{V}_{k1} = \frac{i v_k}{v},
\label{nvkxclx}
\ee
where $v^2 \equiv \sum_{k=1}^{n_d} |v_k|^2$
($v$ is real and positive by definition).

There is an $n \times n$ matrix
\be
\tilde{\mathcal{U}} =
\left( \begin{array}{c} \mathcal{U} \\ \mathcal{T} \end{array} \right)
\label{tildeU}
\ee
that is unitary,
implying that
\be
\mathcal{U} \mathcal{U}^\dagger = \mathbbm{1}_{n_d \times n_d}.
\label{jvkcpf}
\ee
The matrix $\mathcal{T}$ in Eq.~\eqref{tildeU} only exists
when the number $n_c$ of charged scalar $SU(2)_L$ singlets is nonzero.
There is an $m \times m$ matrix
\be
\tilde{\mathcal{V}} =
\left( \begin{array}{c} \mathrm{Re}\, \mathcal{V} \\
  \mathrm{Im}\, \mathcal{V} \\
  \mathcal{R} \end{array} \right)
\label{jdisos}
\ee
that is real and orthogonal.
Therefore,
\bs
\label{vtilde}
\ba
\textrm{Re}\, \mathcal{V}\ \textrm{Re}\, \mathcal{V}^T
&=& \mathbbm{1}_{n_d \times n_d}, \label{1} \\
\textrm{Im}\, \mathcal{V}\ \textrm{Im}\, \mathcal{V}^T
&=& \mathbbm{1}_{n_d \times n_d}, \label{2} \\
\textrm{Re}\, \mathcal{V}\ \textrm{Im}\, \mathcal{V}^T
&=& 0_{n_d \times n_d}, \label{3} \\
\textrm{Im}\, \mathcal{V}\ \textrm{Re}\, \mathcal{V}^T
&=& 0_{n_d \times n_d}.
\label{eq:187}
\ea
\es
The matrix $\mathcal{R}$ in Eq.~\eqref{jdisos}
only exists in models with $n_n \neq 0$.

One can show~\cite{Grimus:2007if}
that in this class of models
\ba
X_{a a^\prime} &=&  s_W^2 \delta_{a a^\prime} -
\frac{\left( \mathcal{U}^T \mathcal{U}^\ast \right)_{a a^\prime}}{2},
\label{Xaa} \\ &=&
\frac{s_W^2 - c_W^2}{2}\, \delta_{a a^\prime}
+ \frac{\left( \mathcal{T}^T \mathcal{T}^\ast \right)_{a a^\prime}}{2},
\label{Xaa2} \\ 
Y_{l l^\prime} &=&  -
\frac{1}{4}\,
\textrm{Im} \left( \mathcal{V}^\dagger \mathcal{V} \right)_{l l^\prime}.
\label{XY}
\ea
Moreover,
\be
y_l = - \textrm{Im} \left( \mathcal{V}^\dagger \mathcal{V} \right)_{1l},
\ee
leading to $y_{l=1} = 0$,
because $\mathcal{V}^\dagger \mathcal{V}$ is Hermitian and therefore
$\textrm{Im} \left( \mathcal{V}^\dagger \mathcal{V} \right)_{11} = 0$.
Thus,
the sum in Eq.~\eqref{ZZS}
really starts at $l=2$,
\textit{viz.}\ there is no vertex $Z Z G^0$,
just as there is no vertex $Z Z Z$.

\subsection{Cancellation of the divergences}

It follows from Eqs.~\eqref{Htb},
\eqref{Sbb},
and~\eqref{jcksps}--\eqref{def:U&V} that
\ba
\label{h1}
c_a &=& \sum_{k=1}^{n_d} \mathcal{U}_{ka}^\ast e_k = \left( \mathcal{U}^\dagger E \right)_a, \\[3mm]
\label{h2}
d_a &=& \sum_{k=1}^{n_d} \mathcal{U}_{ka} f_k = \left( \mathcal{U}^T F \right)_a,\\[3mm]
\label{rels}
r_l &=& - \frac{1}{\sqrt{2}} \sum_{k=1}^{n_d} \mathcal{V}_{kl} f_k = - \frac{1}{\sqrt{2}} \left( \mathcal{V}^T F \right)_l,
\ea
where we have defined the $n_d \times 1$ vectors
\begin{equation}
E = \left( \begin{array}{c} e_1 \\ e_2 \\ \vdots \\ e_{n_d} \end{array} \right),
\quad \quad
F = \left( \begin{array}{c} f_1 \\ f_2 \\ \vdots \\ f_{n_d} \end{array} \right).
\end{equation}

From Eqs.~\eqref{nvkxclx} and~\eqref{h1}--\eqref{rels},
\bs
\ba
\left| c_1 \right| &=& \left| \sum_{k=1}^{n_d} \frac{v_k^\ast}{v}\, e_k \right|
=
\frac{\sqrt{2} m_t}{v},
\label{mdksls} \\
\left| d_1 \right| &=& \left| \sum_{k=1}^{n_d} \frac{v_k}{v}\, f_k \right|
=
\frac{\sqrt{2} m_b}{v} \equiv 0,
\\
\left| r_1 \right| &=&
\left| \frac{1}{\sqrt{2}}\, \frac{v_k}{v}\, f_k \right|
=
\frac{m_b}{v} \equiv 0.
\ea
\es

We further define the $m \times 1$ column vector
\be
R = \left( \begin{array}{c} r_1 \\ r_2 \\ \vdots \\ r_m \end{array} \right).
\ee
It then follows from Eq.~\eqref{rels} that
\ba
\sum_{l=1}^m \left| r_l \right|^2
&=& \frac{1}{2}\, F^T \mathcal{V} \mathcal{V}^\dagger F^\ast
\no &=& \frac{1}{2}\, F^T \left( \mathrm{Re}{\mathcal{V}}
+ i\, \mathrm{Im}{\mathcal{V}} \right)
\left( \mathrm{Re}{\mathcal{V}}^T
- i\, \mathrm{Im}{\mathcal{V}}^T \right) F^\ast
\no &=& \frac{1}{2}\, F^T \Big( \mathrm{Re}{\mathcal{V}}\,
\mathrm{Re}{\mathcal{V}}^T + \mathrm{Im}{\mathcal{V}}\,
\mathrm{Im}{\mathcal{V}}^T + i\, \mathrm{Im}{\mathcal{V}}\, \mathrm{Re}{\mathcal{V}}^T
- i\, \mathrm{Re}{\mathcal{V}}\, \mathrm{Im}{\mathcal{V}}^T \Big) F^\ast. \no
\ea
We now use Eqs.~\eqref{vtilde} to obtain
\ba
\sum_{l=1}^m \left| r_l \right|^2 &=&
\frac{1}{2}\, F^T \Big( \mathbbm{1}_{n_d \times n_d} + \mathbbm{1}_{n_d \times n_d}
+ i \times 0_{n_d \times n_d} - i \times 0_{n_d \times n_d} \Big) F^\ast
\no &=& F^T F^\ast = \sum_{k=1}^{n_d} \left| f_k \right|^2.
\label{jgkvpdd}
\ea

From Eqs.~\eqref{h1} and~\eqref{jvkcpf},
\be
\sum_{a=1}^{n} |c_a|^2 =
E^\dagger \mathcal{U} \mathcal{U}^\dagger E = E^\dagger E = \sum_{k=1}^{n_d} |e_k|^2.
\label{cack}
\ee
Notice that the two sums in Eq.~\eqref{cack} run over different spaces
(up to $n$ and $n_d$, respectively).  Similarly,
\be
\sum_{a=1}^{n} |d_a|^2 = \sum_{k=1}^{n_d} |f_k|^2.
\label{jhsssi}
\ee

From Eqs.~\eqref{Xaa},
\eqref{h1},
and~\eqref{jvkcpf},
\begin{eqnarray}
\sum_{a, a' = 1}^n c_{a} X_{a a'} c_{a'}^\ast \,
&=& s_W^2\, E^T \mathcal{U}^\ast \mathcal{U}^T E^\ast
- \frac{E^T \mathcal{U}^\ast \mathcal{U}^T \mathcal{U}^\ast \mathcal{U}^T
  E^\ast}{2}
\no
&=& s_W^2\, E^T E^\ast - \frac{E^T E^\ast}{2}
\no
&=& \frac{s_W^2 - c_W^2}{2} \sum_{k=1}^{n_d} \left| e_k \right|^2
\no
&=& \frac{s_W^2 - c_W^2}{2} \sum_{a=1}^{n} \left| c_a \right|^2,
\end{eqnarray}
where the last equality follows from Eq.~\eqref{cack}.
This proves that this class of models
obeys the consistency Eq.~\eqref{eq:165a}.
Similarly,
one can show that Eq.~\eqref{eq:165b} is also obeyed,
confirming within this class of models
the cancellation of the divergences of the contributions from charged scalars.

Next we compute 
\ba
\sum_{l, l^\prime = 1}^m r_l\,
\mathrm{Im}{\left( \mathcal{V}^\dagger \mathcal{V} \right)_{l l^\prime}}\,
r_{l^\prime}^\ast
&=& \frac{1}{2}\, F^T \mathcal{V}\
\mathrm{Im}\Big[
    \left( \mathrm{Re}{\mathcal{V}}^T - i\, \mathrm{Im}{\mathcal{V}}^T \right)
    \left( \mathrm{Re}{\mathcal{V}} + i\, \mathrm{Im}{\mathcal{V}} \right)
    \Big]
\mathcal{V}^\dagger F^\ast
\no &=& \frac{1}{2}\, F^T
\left( \mathrm{Re}{\mathcal{V}} + i\, \mathrm{Im}{\mathcal{V}} \right)
\left( \mathrm{Re}{\mathcal{V}}^T\, \mathrm{Im}{\mathcal{V}}
- \mathrm{Im}{\mathcal{V}}^T\, \mathrm{Re}{\mathcal{V}} \right)
\no & & \qquad \times
\left( \mathrm{Re}{\mathcal{V}}^T - i\, \mathrm{Im}{\mathcal{V}}^T \right)
F^\ast.
\ea
We use once again Eqs.~\eqref{vtilde} to obtain
\ba
\sum_{l, l^\prime = 1}^m r_l\,
\mathrm{Im}{\left( \mathcal{V}^\dagger \mathcal{V} \right)_{l l^\prime}}\,
r_{l^\prime}^\ast &=&
\frac{1}{2}\, F^T
\left( \mathrm{Im}{\mathcal{V}} - i\, \mathrm{Re}{\mathcal{V}} \right)
\times \left( \mathrm{Re}{\mathcal{V}}^T - i\, \mathrm{Im}{\mathcal{V}}^T \right)
F^\ast
\no &=& \frac{1}{2}\, F^T
\left( - 2 i \times \mathbbm{1}_{n_d \times n_d} \right) F^\ast
\no &=& - i\, F^T F^\ast
\no &=& - i\, \sum_{k=1}^{n_d} \left| f_k \right|^2
\no &=& - i\, \sum_{l=1}^{m} \left| r_l \right|^2,
\ea
where in the last step we have used Eq.~\eqref{jgkvpdd}.
Taking into account Eq.~\eqref{XY},
we conclude that in this class of models
the consistency Eq.~\eqref{eq:consistent_N} also holds.

\subsection{Simplification of the charged-scalars contribution}

In this class of models,
from Eqs.~\eqref{Xaa2} and~\eqref{difference},
\be
\begin{split}
X_{a a^\prime}
&= \left( g_{Lb}^0 - g_{Rt}^0 \right) \delta_{a a^\prime}
+ \frac{\left( \mathcal{T}^T \mathcal{T}^\ast \right)_{a a^\prime}}{2}
\no 
&= \left( g_{Rb}^0 - g_{Lt}^0 \right) \delta_{a a^\prime}
+ \frac{\left( \mathcal{T}^T \mathcal{T}^\ast \right)_{a a^\prime}}{2}.
\end{split}
\ee
Therefore,
one may write the charged-scalars contribution as
\ba
&&\left( 16 \pi^2 \right) \Delta g_{Lb} =
\sum_{a=1}^n \left| c_a \right|^2 \left\{ {\vbox to 18 pt{}} - g_{Lt}^0 m_t^2 C_0 \left( 0, M_Z^2, 0, m_a^2, m_t^2, m_t^2 \right)
\right. \no & &
+ g_{Rt}^0 \bigg[ 2\, C_{00} \left( 0, M_Z^2, 0, m_a^2, m_t^2, m_t^2 \right) 
- \frac{1}{2} - 2\, C_{00} \left( 0, M_Z^2, 0, m_t^2, m_a^2, m_a^2 \right)
  \no & &
  \hspace{10ex}
  - M_Z^2\, C_{12} \left( 0, M_Z^2, 0, m_a^2, m_t^2, m_t^2 \right) \bigg]
\no & & \left.
+ g_{Lb}^0 \Big[ B_1 \left( 0, m_t^2, m_a^2 \right)
  + 2\, C_{00} \left( 0, M_Z^2, 0, m_t^2, m_a^2, m_a^2 \right) \Big]
{\vbox to 18 pt{}} \right\}
\no & &
+ \! \! \sum_{a, a^\prime = 1}^n \left( \mathcal{T}^T \mathcal{T}^\ast \right)_{a a^\prime}
c_a c_{a^\prime}^\ast\,
C_{00} \left( 0, M_Z^2, 0, m_t^2, m_{a^\prime}^2, m_a^2 \right).
\label{simple_DgL_H}
\ea
The first column of the matrix $\mathcal{T}$ is zero,
because
$\sum_k \left| \mathcal{U}_{k1} \right|^2
= \left. \sum_k \left| v_k \right|^2 \right/ v^2 = 1$.
Thus,
$\left( \mathcal{T}^T \mathcal{T}^\ast \right)_{1a}
= \left( \mathcal{T}^T \mathcal{T}^\ast \right)_{a1} = 0$
and the charged Goldstone boson does not contribute to the sum in the
last line of Eq.~\eqref{simple_DgL_H}.  On the other hand, the
Goldstone boson does contribute to the sum over $a$ in the first five lines,
but $\left| c_1 \right|$ has the same value as in the SM,
\textit{cf.}\ Eq.~\eqref{mdksls};
therefore,
the contribution of the charged Goldstone boson
{is the same as in the SM}
and should be subtracted out.
The simplified expression for the charged-scalar contributions
to $\Delta g_{Rb}$ is obtained from Eq.~\eqref{simple_DgL_H} through
the changes $c_a \to d_a^\ast$ and $L \leftrightarrow R$.

Suppose a model with no charged $SU(2)_L$ singlets.  Then the matrix
$\mathcal{T}$ does not exist.  If one furthermore makes the
approximation $M_Z = 0$, then the contribution of the charged scalars
in Eq.~\eqref{simple_DgL_H} becomes
\ba
\left( 16 \pi^2 \right) \Delta g_{Lb} &=&
\sum_{a=1}^n \left| c_a \right|^2 \left\{
- g_{Lt}^0 m_t^2 C_0 \left( 0, 0, 0, m_a^2, m_t^2, m_t^2 \right)
\right.
\no & &
+ 2 g_{Rt}^0 \left[ C_{00} \left( 0, 0, 0, m_a^2, m_t^2, m_t^2 \right)
  - C_{00} \left( 0, 0, 0, m_t^2, m_a^2, m_a^2 \right)
  \right. \no & & \left.
  - \frac{1}{4} \right]
\no & & \left.
+ g_{Lb}^0 \left[ B_1 \left( 0, m_t^2, m_a^2 \right)
  + 2\, C_{00} \left( 0, 0, 0, m_t^2, m_a^2, m_a^2 \right) \right]
{\vbox to 18 pt{}} \right\},
\ea
and similarly for $\Delta g_{Rb}$,
with $c_a \rightarrow d_a$ and $L \leftrightarrow R$.
One easily finds that
\be
B_1 \left( 0, m_t^2, m_a^2 \right)
+ 2\, C_{00} \left( 0, 0, 0, m_t^2, m_a^2, m_a^2 \right) = 0,
\ee
and that
\ba
& & C_{00} \left( 0, 0, 0, m_a^2, m_t^2, m_t^2 \right)
- C_{00} \left( 0, 0, 0, m_t^2, m_a^2, m_a^2 \right) - \frac{1}{4}
\no
&=& \frac{m_t^2}{2}\, C_0 \left( 0, 0, 0, m_a^2, m_t^2, m_t^2 \right).
\ea
Hence,
\bs
\label{finalr}
\ba
\left( 16 \pi^2 \right)
\Delta g_{Lb} &=& \sum_{a=1}^n \left| c_a \right|^2
\left( g_{Rt}^0 - g_{Lt}^0  \right)
m_t^2\, C_0 \left( 0, 0, 0, m_a^2, m_t^2, m_t^2 \right)
\no &=&
- \sum_{a=1}^n \frac{\left| c_a \right|^2}{2}\,
m_t^2\, C_0 \left( 0, 0, 0, m_a^2, m_t^2, m_t^2 \right),
\\
\left( 16 \pi^2 \right)
\Delta g_{Rb} &=& \sum_{a=1}^n \left| d_a \right|^2
\left( g_{Lt}^0 - g_{Rt}^0  \right)
m_t^2\, C_0 \left( 0, 0, 0, m_a^2, m_t^2, m_t^2 \right)
\no &=&
+ \sum_{a=1}^n \frac{\left| d_a \right|^2}{2}\,
m_t^2\, C_0 \left( 0, 0, 0, m_a^2, m_t^2, m_t^2 \right).
\ea
\es
The dependence on $\theta_W$ disappeared!  This must indeed happen
because, in the limit $M_Z = 0$, the $Z$ gauge boson is
indistinguishable from the photon---since they are both massless---,
and therefore the Weinberg angle loses its meaning and must disappear
from any physically meaningful quantity.  The function
\be
m_t^2\, C_0 \left( 0, 0, 0, m_a^2, m_t^2, m_t^2 \right)
= \frac{x}{1 - x} \left( 1 + \frac{\ln{x}}{1 - x} \right), \quad \mbox{with} \ x = \frac{m_t^2}{m_a^2}
\ee
has been given in Eq.~(4.5) of Ref.~\cite{Haber:1999zh} and has been
used in all the subsequent analyses, by many authors, of models with
extra doublets (and possibly neutral singlets).  In our more general
result~\eqref{simple_DgL_H}, though, we keep CP violation,
{we allow for charged singlets and we do not make $M_Z = 0$.}

As a consequence of Eqs.~\eqref{finalr}, in a 2HDM, where there is
only one physical charged scalar,
\be
\frac{\Delta g_{Lb}}{\left| c_2 \right|^2}
= - \frac{\Delta g_{Rb}}{\left| d_2 \right|^2}.
\label{uvidosp}
\ee
In general, as long as there are no charged singlets and the
approximation $M_Z \approx 0$ is good, $\Delta g_{Lb}$ and $\Delta
g_{Rb}$ have opposite signs when the contribution of the neutral
scalars is not taken into account.

\section{\label{sec:exp}Connection with experiment}

The couplings $g_{Lb}$ and $g_{Rb}$ in Eq.~\eqref{Zbb}
may be determined experimentally from:\footnote{See the discussion
  by Erler and Freitas in Ref.~\cite{Tanabashi:2018oca}.}
\begin{enumerate}
\item The rate
\be
R_b =
\frac{\Gamma \left( Z \rightarrow b \bar{b} \right)}{\Gamma
\left( Z \rightarrow \textrm{hadrons} \right)}.
\label{R_b}
\ee
\item Several asymmetries,
including
\begin{enumerate}
%
\item the $Z$-pole forward--backward asymmetry measured at LEP1
\be
A_{FB}^{(0,f)}
=
\frac{\sigma \left( e^- \rightarrow b_F \right)
- \sigma \left( e^- \rightarrow b_B \right)}{\sigma \left( e^-
\rightarrow b_F \right) + \sigma \left( e^- \rightarrow b_B \right)}
= \frac{3}{4} A_e  A_b,
\label{A_FB}
\ee
where $b_F$ ($b_B$) stands for final-state bottom quarks
moving in the forward (backward) direction
with respect to the direction of the initial-state electron;
\item the left--right forward--backward asymmetry
  measured by the SLD Collaboration
\be
\begin{split}
A_{LR}^{FB} \left( b \right) &=  \frac{\sigma \! \left(e_L^- \! \rightarrow \! b_F \right)
  \! - \sigma \left(e_L^- \! \rightarrow \! b_B \right)
  \! - \sigma \left(e_R^- \! \rightarrow \! b_F \right)
  \! + \sigma \left(e_R^- \! \rightarrow \! b_B \right)}{
  \sigma \left(e_L^- \! \rightarrow \! b_F \right)
  \! + \sigma \left(e_L^- \! \rightarrow \! b_B \right)
  \! + \sigma \left(e_R^- \! \rightarrow \! b_F \right)
  \! + \sigma \left(e_R^- \! \rightarrow \! b_B \right)} \\
& = \frac{3}{4} A_b,
\label{A_LRFB}
\end{split}
\ee
where $e_L^-$ ($e_R^-$) are initial-state left-handed (right-handed) electrons.
\end{enumerate}
\end{enumerate}
Introducing the vector- and axial-vector bottom-quark couplings
\be
v_b = g_{Lb} + g_{Rb},
\quad \quad
a_b = g_{Lb} - g_{Rb},
\quad \quad
\mbox{and} \quad \quad r_b = \frac{v_b}{a_b},
\label{a_and_b}
\ee
one has~\cite{Haber:1999zh, Field:1997gz}
\ba
R_b &=&
\left( 1 + \frac{\Sigma}{s_b\, \eta^\textrm{QCD}\, \eta^\textrm{QED}} \right)^{-1},
\label{uidosp}
\\
A_b &=&
\frac{2\, r_b\, \sqrt{1 - 4 \mu_b}}{1 - 4 \mu_b
  + \left( 1 + 2 \mu_b \right) r_b^2}.
\ea
In Eq.~\eqref{uidosp},
$\eta^\textrm{QCD}=0.9953$ and $\eta^\textrm{QED}=0.99975$
are QCD and QED corrections,
respectively.
Moreover,
\ba
\mu_b &=& \frac{m_b \left( M_Z \right)^2}{M_Z^2},
\\
s_b &=& \left( 1 - 6 \mu_b \right) a_b^2 + v_b^2,
\\
\Sigma &=& \sum_{q=u,d,s,c} \left( a_q^2 + v_q^2 \right).
\ea
Neglecting $\mu_b \approx 10^{-3}$
and setting the QCD and QED corrections to unity,
one gets
\ba
R_b & \approx &
\frac{2 \left( g_{Lb}^2 + g_{Rb}^2 \right)}{2 \left( g_{Lb}^2 + g_{Rb}^2 \right)
  + \Sigma},
\\
A_b & \approx & \frac{g_{Lb}^2 - g_{Rb}^2}{g_{Lb}^2 + g_{Rb}^2}.
\label{vjoidps}
\ea
Equation~\eqref{vjoidps} with $b \rightarrow e$
defines the $A_e$ appearing in Eq.~\eqref{A_FB},
which has also been determined experimentally.

The recent fit to the electroweak data
by Erler and Freitas in Ref.~\cite{Tanabashi:2018oca} finds
\bs
\label{uvido}
\ba
R_b^\textrm{fit} &=& 0.21629 \pm 0.00066,
\\
A_b^\textrm{fit} &=& 0.923 \pm 0.020,
\ea
\es
to be compared with the SM values $R_b^\textrm{SM} = 0.21582 \pm
0.00002$ and $A_b^\textrm{SM} = 0.9347$.  Thus, the experimental $R_b$
is about $0.7\sigma$ above the SM value, while $A_b$ is about
$0.6\sigma$ below the SM value.  However, this good agreement only
applies to the overall fit of many observables producing
Eqs.~\eqref{uvido}.  The measured values of the bottom-quark
asymmetries by themselves alone reveal a much larger discrepancy; as
pointed out in Ref.~\cite{Tanabashi:2018oca}, extracting $A_b$ from
$A_{FB}^{(0,b)}$ and using $A_e = 0.1501 \pm 0.0016$ leads to a result
which is $3.1 \sigma$ below the SM (the precise value of $A_b$ depends
on which observables $A_e$ is extracted from), while combining
$A_{FB}^{(0,b)}$ with $A_{LR}^{FB}$ leads to $A_b = 0.899 \pm 0.013$,
which deviates from the SM value by $2.8 \sigma$.

There are, thus, two possible approaches.  The first one consists in
taking as good the values~\eqref{uvido} obtained from the SM fit and
using $R_b^\textrm{fit}$ and $A_b^\textrm{fit}$ as constraints on New
Physics (NP).  The second one is seeking NP that might explain an
$R_b$ just slightly above the SM, together with an $A_b$ that
undershoots the SM by $2.8 \sigma$.

It is convenient to switch from the parameterization in
Eq.~\eqref{eq:57}, which splits the couplings $g_{\aleph b}$ as
$g_{\aleph b}^0 + \Delta g_{\aleph b}$, where $g_{\aleph b}^0$ is the
tree-level piece and $\Delta g_{\aleph b}$ is the one-loop piece, to
the alternative parameterization
\be
g_{\aleph b} = g_{\aleph b}^\textrm{SM} + \delta g_{\aleph b},
\label{delta_gLR}
\ee
which splits them into the SM piece $g_{\aleph b}^\textrm{SM}$ (which
includes the SM loop correction) and the NP piece $\delta g_{\aleph
b}$.  A simple rule of thumb can be obtained by expanding to first
order in the deviations; one finds~\cite{Haber:1999zh}
\bs
\label{cjvidop}
\ba
\delta R_b &=& - 0.7785\, \delta g_{L b} + 0.1409\, \delta g_{R b},
\label{eq:75a}
\\
\delta A_b &=& - 0.2984\, \delta g_{L b} - 1.6234\, \delta g_{R b}.
\label{eq:75b}
\ea
\es
This shows that,
assuming (rather arbitrarily) $\delta g_{R b} \approx - \delta g_{L b}$,
$\delta R_b$ is pulled down (up)
and $\delta A_b$ is pulled up (down)
by a positive (negative) $\delta g_{L b}$.
Inverting Eqs.~\eqref{cjvidop}~\cite{Haber:1999zh},
\bs
\ba
\delta g_{L b} &=& - 1.2433\, \delta R_b - 0.1079\, \delta A_b,
\\
\delta g_{R b} &=& 0.2286 \, \delta R_b - 0.5962\, \delta A_b.
\ea
\es
If one wishes to follow the second approach above,
\textit{viz.}\ using NP to keep $R_b$ close to its SM value
while reducing $A_b$ significantly,
then one needs to get a small $\delta g_{L b}$
together with a significant \textit{positive} $\delta g_{R b}$.

\section{\label{sec:particular}Simple particular cases}

\subsection{The 2HDM in an alignment limit}

In the 2HDM, one may always employ the `Higgs basis' for the scalar
doublets $\Phi_{1,2}$; in that basis,
\be
\label{nvjfio}
\Phi_1 = \left( \begin{array}{c}
G^+ \\ \left( v + \rho_1 + i G^0 \right) \left/ \sqrt{2} \right.
\end{array} \right),
\quad 
\Phi_2 = \left( \begin{array}{c}
H^+ \\ \left( \rho_2 + i \eta \right) \left/ \sqrt{2} \right.
\end{array} \right),
\ee
where $G^+ = H_1^+$ and $G^0 = S_1^0$ are the Goldstone bosons
and $H^+ = H_2^+$ is a physical charged scalar.
Then,
the Yukawa couplings $e_1$ and $f_1$ are simply
\begin{equation}
\label{eq:201}
e_1 = \frac{\sqrt{2} m_t}{v},
\quad \quad
f_1 = \frac{\sqrt{2} m_b}{v},
\end{equation}
which may be taken to be real and positive.
In this section we shall \emph{assume} that,
for some unspecified reason,
the neutral fields $\rho_{1,2}$ and $\eta$ in Eq.~\eqref{nvjfio}
coincide with the physical neutral scalars,
\textit{viz.}\ $S_2^0 = \rho_1$,
$S_3^0 = \rho_2$,
and $S_4^0 = \eta$.
We moreover assume that $S_2^0$ is the scalar particle discovered at the LHC,
with mass $m_2 = 125$\,GeV.
That means,
we assume
an `alignment limit'~\cite{Gunion:2002zf}
of the 2HDM wherein $S_2^0$ couples to the gauge bosons
and to the top and bottom quarks
with exactly the same strength as the Higgs boson of the SM.
The matrix $\mathcal{U}$ defined by Eq.~\eqref{jdkpsos}
and the matrix $\mathcal{V}$ defined by Eq.~\eqref{def:U&V} are
\begin{equation}
\label{eq:200}
\mathcal{U} = \left( \begin{array}{cc} 1 & 0 \\ 0 & 1 \end{array} \right),
\quad \quad
\mathcal{V} = \left( \begin{array}{cccc} i & 1 & 0 & 0 \\ 0 & 0 & 1 & i
\end{array} \right).
\end{equation}
Since $\mathcal{U}$ is,
in this case,
the  $2 \times 2$ unit matrix, we have, from Eqs.~\eqref{h1} and~\eqref{h2},
$c_1 = e_1$,
$c_2 = e_2$,
$d_1 = f_1$,
and $d_2 = f_2$.
The free parameters in our model are the mass $M_{H^{+}}$ of the
charged scalar, the masses  $m_3$ and $m_4$ of the two
new neutral scalars $S_3^0$ and $S_4^0$, respectively, and the Yukawa
couplings $c_2$ and $d_2$.\footnote{The 2HDM in this subsection is not
endowed with the usual $\mathbbm{Z}_2$ symmetry that prevents the
appearance of flavor-changing neutral currents (FCNC). Therefore, a
multi-generation version of this (toy) model will in general be
plagued by FCNC and by the need for their suppression.  This needs not
concern us here, since we are considering a truncated version of the
model only with the third generation.}
Since there are no charged singlets,
\begin{equation}
  \label{eq:202}
  X_{aa'}= \frac{s_W^2 - c_W^2}{2}\, \delta_{aa'},
\end{equation}
while,
from the matrix $\mathcal{V}$ in Eq.~\eqref{eq:200},
\bs
\ba
\label{eq:203}
\textrm{Im} \left( \mathcal{V}^\dagger \mathcal{V} \right) &=&
\left( \begin{array}{cccc} 0 & -1 & 0 & 0 \\ 1 & 0 & 0 & 0 \\
  0 & 0 & 0 & 1 \\ 0 & 0 & -1 & 0 \end{array} \right),
\\
R = \left( \begin{array}{c} r_1 \\ r_2 \\ r_3 \\ r_4 \end{array}
\right) &=& -\frac{1}{\sqrt{2}} \left( \begin{array}{c}
  i\, d_1 \\ d_1 \\ d_2 \\ i\, d_2 \end{array} \right).
\ea
\es

\subsubsection{Charged-scalar contribution}

\begin{figure}[htb]
\centering
\includegraphics[width=0.65\textwidth]{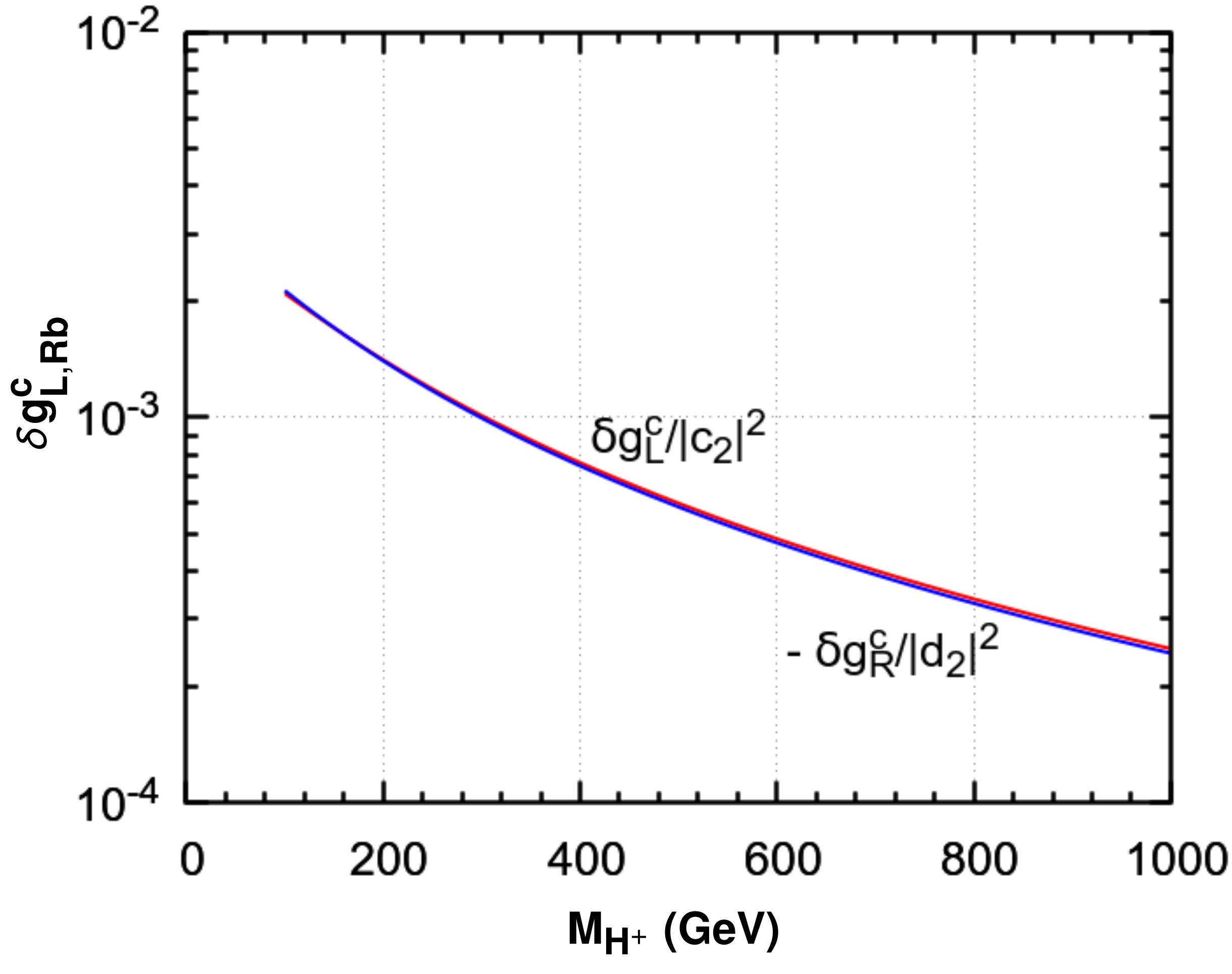}
\caption{Contribution of the charged scalar
  to $\delta g_{Lb}$ (red curve)
  and to $- \delta g_{Rb}$ (blue curve)
  in a general 2HDM.}
\label{fig:2hdm-charged}
\end{figure}

\begin{figure}[htb]
\centering
\includegraphics[width=0.65\textwidth]{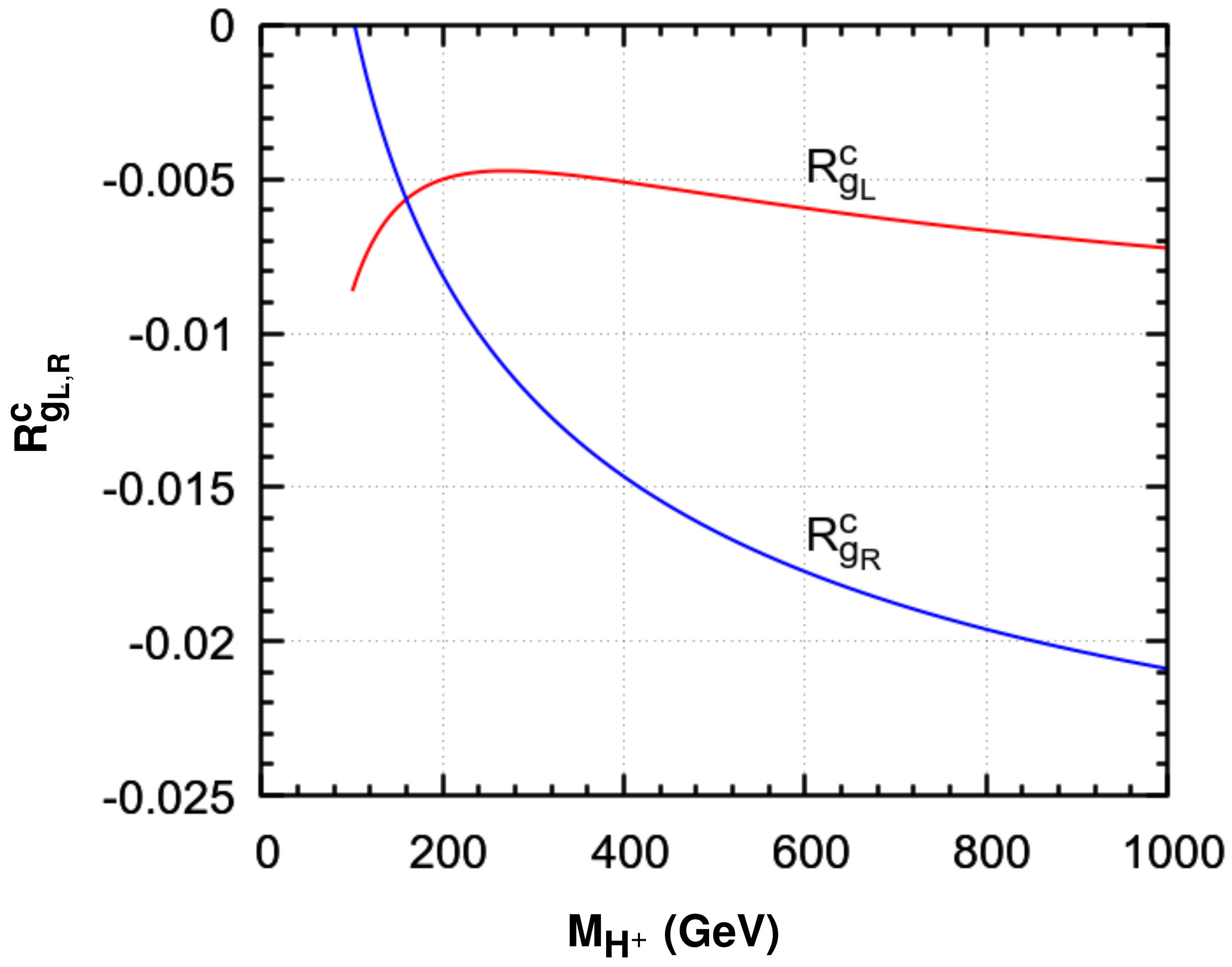}
\caption{In red: the asymmetry between
  $\delta g_{Lb}^c$ computed using $M_Z =91$\,GeV
  and the same quantity computed using $M_Z = 0$.
  In blue: the asymmetry between
  $\delta g_{Rb}^c$ computed with $M_Z =91$\,GeV
  and the same quantity computed with $M_Z = 0$.}
  \label{fig:2hdm-charged-assymetry}
\end{figure}

Let us denote by superscripts $c$ and $n$ the new-physics
contributions to $\delta g_{Lb}$ and $\delta g_{Rb}$ coming from the
charged and neutral scalars, respectively.  In the charged-scalar
sector of a generic 2HDM, the contribution of the charged Goldstone
boson can be separated and included in the SM.  The genuine new
contribution is
\bs
\label{eq:99999}
\ba
\label{eq:9}
\delta g_{Lb}^c &=&
\frac{\left| c_2 \right|^2}{16\pi^2} \left\{ \left( s_W^2 - c_W^2 \right)
  C_{00} \left( 0, M_Z^2, 0, m_t^2, M_{H^{+}}^2, M_{H^{+}}^2 \right)
  \right. \no & &
  - g_{Lt}^0\, m_t^2\, C_{0} \left( 0, M_Z^2, 0, M_{H^{+}}^2, m_t^2, m_t^2 \right)
  \no & &
  + g_{Rt}^0 \left[ 2\, C_{00} \left( 0, M_Z^2, 0, M_{H^{+}}^2, m_t^2, m_t^2 \right)
  - \frac{1}{2}
  \right. \no & & \left. \left.
  - M_Z^2\, C_{12} \left( 0, M_Z^2, 0, M_{H^{+}}^2, m_t^2, m_t^2 \right)
  \right]
  + g_{Lb}^0\, B_1 \left( 0, m_t^2, M_{H^{+}}^2 \right) \right\},
  \hspace*{7mm}
\\*[2mm]
\delta g_{Rb}^c &=&
\frac{\left| d_2 \right|^2}{16\pi^2} \left\{
  \left( s_W^2 - c_W^2 \right)
  C_{00} \left( 0, M_Z^2, 0, m_t^2, M_{H^{+}}^2, M_{H^{+}}^2 \right)
  \right. \no & &
  - g_{Rt}^0\, m_t^2\, C_{0} \left( 0, M_Z^2, 0, M_{H^{+}}^2, m_t^2,
    m_t^2 \right) 
  \no & &
  + g_{Lt}^0 \left[ 2\, C_{00} \left( 0, M_Z^2, 0, M_{H^{+}}^2, m_t^2,
      m_t^2 \right) 
    - \frac{1}{2}
    \right. \no & & \left. \left.
    - M_Z^2\, C_{12} \left( 0, M_Z^2, 0, M_{H^{+}}^2, m_t^2, m_t^2 \right)
  \right] + g_{Rb}^0\, B_1 \left( 0, m_t^2, M_{H^{+}}^2\right) \right\}.
\label{eq:9b}
\ea
\es
If we plot $\delta g_{Lb}^c \left / \left| c_2 \right|^2 \right.$
and $\delta g_{Rb}^c \left/ \left| d_2 \right|^2 \right.$, we get
general results for any 2HDM.  We have used
\texttt{LoopTools}~\cite{Hahn:1998yk} to perform the numerical
integrations contained in the Passarino--Veltman functions.  The
results are shown in Fig.~\ref{fig:2hdm-charged}.
One sees that $0 < \delta g_{Lb}^c \lesssim 0.002 \left| c_2 \right|^2$
and that Eq.~\eqref{uvidosp} holds to an excellent approximation;
this indicates that the approximation $M_Z = 0$ is in fact very good.
This is vindicated by Fig.~\ref{fig:2hdm-charged-assymetry},
which displays the asymmetries $R_{g_{L,R}}$
between the values of $\delta g_{Lb}^c \left/ \left| c_2 \right|^2 \right.$
and $\delta g_{Rb}^c \left/ \left| d_2 \right|^2 \right.$
computed with $M_Z \neq 0$ and with $M_Z = 0$:
\bs
\label{eq:1}
\ba
R_{g_{L}}^c &=& \frac{\delta g_{Lb}^c \left( M_Z \right) -
  \delta g_{Lb}^c \left( 0 \right)}{\delta
  g_{Lb}^c \left( M_Z \right) + \delta g_{Lb}^c \left( 0 \right)},
\\
R_{g_{R}}^c &=&
R_{g_{L}}^c \left( L \to R \right).
\ea
\es
One observes in Fig.~\ref{fig:2hdm-charged-assymetry}
that both asymmetries are at most of order 1\%.

\subsubsection{Neutral-scalar contribution}

Taking into account Eq.~\eqref{XY}, in the class of models of
Section~\ref{sec:doubsing}, Eq.~\eqref{eq:153N} reads
\be
\Delta g_{Lb} \left( a \right) = \frac{-i}{16 \pi^2} \sum_{l,l'=1}^{m} r_l\,
\mathrm{Im} \left( \mathcal{V}^\dagger \mathcal{V} \right)_{ll^\prime}
r_{l^\prime}^\ast\,
C_{00} \left( 0, M_Z^2, 0, 0, m_{l'}^2, m_l^2 \right).
\label{jvidoii}
\ee
Since $
C_{00} \left( 0, M_Z^2, 0, 0, m_{l'}^2, m_l^2 \right)
=
C_{00} \left( 0, M_Z^2, 0, 0, m_l^2, \right. $ $ \left. m_{l'}^2 \right),
$
Eq.~\eqref{jvidoii} may be simplified to
\be
\Delta g_{Lb} \left( a \right) =
\frac{1}{8 \pi^2} \sum_{l=1}^{m-1} \sum_{l^\prime=l+1}^m
  \mathrm{Im} \left( \mathcal{V}^\dagger \mathcal{V} \right)_{ll^\prime}\,
  \mathrm{Im} \left( r_l r_{l^\prime}^\ast \right)
  C_{00} \left( 0, M_Z^2, 0, 0, m_{l'}^2, m_l^2 \right).
\label{jvufio}
\ee
%
In the 2HDM of this section, because of Eq.~\eqref{eq:203},
Eq.~\eqref{jvufio} reads
\ba
\Delta g_{Lb} \left( a \right) &=& \frac{1}{16 \pi^2} \left[
  \left| d_1 \right|^2 C_{00} \left( 0, M_Z^2, 0, 0, m_2^2, m_1^2 \right)
  \right. \no & & \left.
- \left| d_2 \right|^2 C_{00} \left( 0, M_Z^2, 0, 0, m_4^2, m_3^2 \right)
\right].
\label{ndlsapa}
\ea
Since $S_1^0 = G^0$ is the neutral Goldstone boson
and $S_2^0$ is the Higgs particle of the SM,
the first term in the right-hand side of Eq.~\eqref{ndlsapa}
is an SM contribution that we are uninterested in;
we just care about the NP contributions,
which are
\bs
\label{mcdlls}
\ba
\delta g_{Lb}^n &=& \frac{\left| d_2 \right|^2}{16 \pi^2} \left\{
- C_{00} \left( 0, M_Z^2, 0, 0, m_3^2, m_4^2 \right)
\right. \nonumber \\
& & + \frac{g_{Rb}^0}{2} \left[
2\, C_{00} \left( 0, M_Z^2, 0, m_3^2, 0, 0 \right)
- \frac{1}{2} - M_Z^2\, C_{12} \left( 0, M_Z^2, 0, m_3^2, 0, 0 \right)
\right. \no & & \left.
+ 2\, C_{00} \left( 0, M_Z^2, 0, m_4^2, 0, 0 \right)
- \frac{1}{2} - M_Z^2\, C_{12} \left( 0, M_Z^2, 0, m_4^2, 0, 0 \right)
\right]
\no & & \left.
+ \frac{g_{Lb}^0}{2} \left[ B_1 \left( 0, 0, m_3^2 \right)
  + B_1 \left( 0, 0, m_4^2 \right) \right] \right\},
\\*[+2mm]
\delta g_{Rb}^n &=& \frac{\left| d_2 \right|^2}{16 \pi^2} \left\{
C_{00} \left( 0, M_Z^2, 0, 0, m_3^2, m_4^2 \right)
\right. \no & &
+ \frac{g_{Lb}^0}{2} \left[
2\, C_{00} \left( 0, M_Z^2, 0, m_3^2, 0, 0 \right)
- \frac{1}{2} - M_Z^2\, C_{12} \left( 0, M_Z^2, 0, m_3^2, 0, 0 \right)
\right. \no & & \left.
+ 2\, C_{00} \left( 0, M_Z^2, 0, m_4^2, 0, 0 \right)
- \frac{1}{2} - M_Z^2\, C_{12} \left( 0, M_Z^2, 0, m_4^2, 0, 0 \right) \right]
\no & & \left.
+ \frac{g_{Rb}^0}{2} \left[ B_1 \left( 0, 0, m_3^2 \right)
  + B_1 \left( 0, 0, m_4^2 \right) \right] \right\}.
\ea
\es

Let us compute the limit $M_Z = 0$ of Eqs.~\eqref{mcdlls}.
Using
\bs
\ba
C_{00} \left( 0, 0, 0, 0, A, B \right) &=&
\frac{\mathrm{div} - \ln{\mu^2}}{4} + \frac{3}{8}
+ \frac{B\, \ln{B} - A\, \ln{A}}{4 \left( A - B \right)},
\\
C_{00} \left( 0, 0, 0, A, 0, 0 \right) - \frac{1}{4} &=&
- \frac{B_1 \left( 0, 0, A \right)}{2},
\\
B_1 \left( 0, 0, A \right) &=&
- \frac{\mathrm{div}}{2} - \frac{1}{4} + \frac{1}{2}\,
\ln{\frac{A}{\mu^2}},
\ea
\es
one obtains the approximation
\be
\label{eq:approx}
\delta g_{Lb}^n \approx - \delta g_{Rb}^n \approx
\frac{\left| d_2 \right|^2}{64 \pi^2}
\left( - 1 + \frac{m_3^2 + m_4^2}{m_3^2 - m_4^2}
  \ln{\frac{m_3}{m_4}} \right),
\ee
which vanishes when $m_3 = m_4$.
One sees that
\begin{itemize}
\item in the limit $M_Z = 0$,
  $\delta g_{Lb}^n = - \delta g_{Rb}^n$;
\item in that limit,
  $\delta g_{Lb}^n$ and $\delta g_{Rb}^n$ are independent of $\theta_W$---this is
  for the reason explained after Eqs.~\eqref{finalr};
\item in that limit,
  $\delta g_{Lb}^n = \delta g_{Rb}^n = 0$
  when the two extra neutral scalars have equal masses.
\end{itemize}

\begin{figure*}[htb]
  \centering
  \begin{tabular}{cc}
    \includegraphics[width=0.47\textwidth]{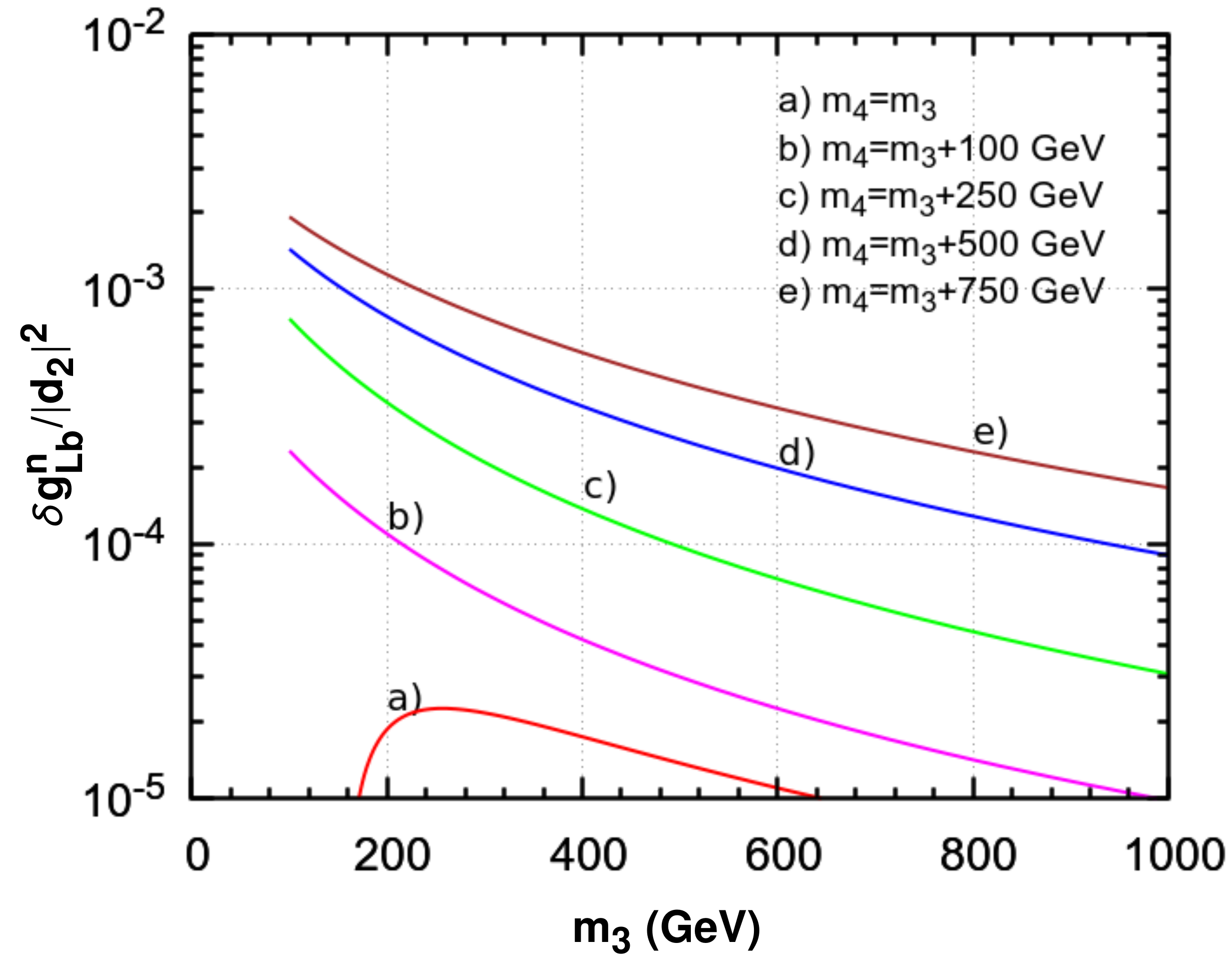}& 
    \includegraphics[width=0.47\textwidth]{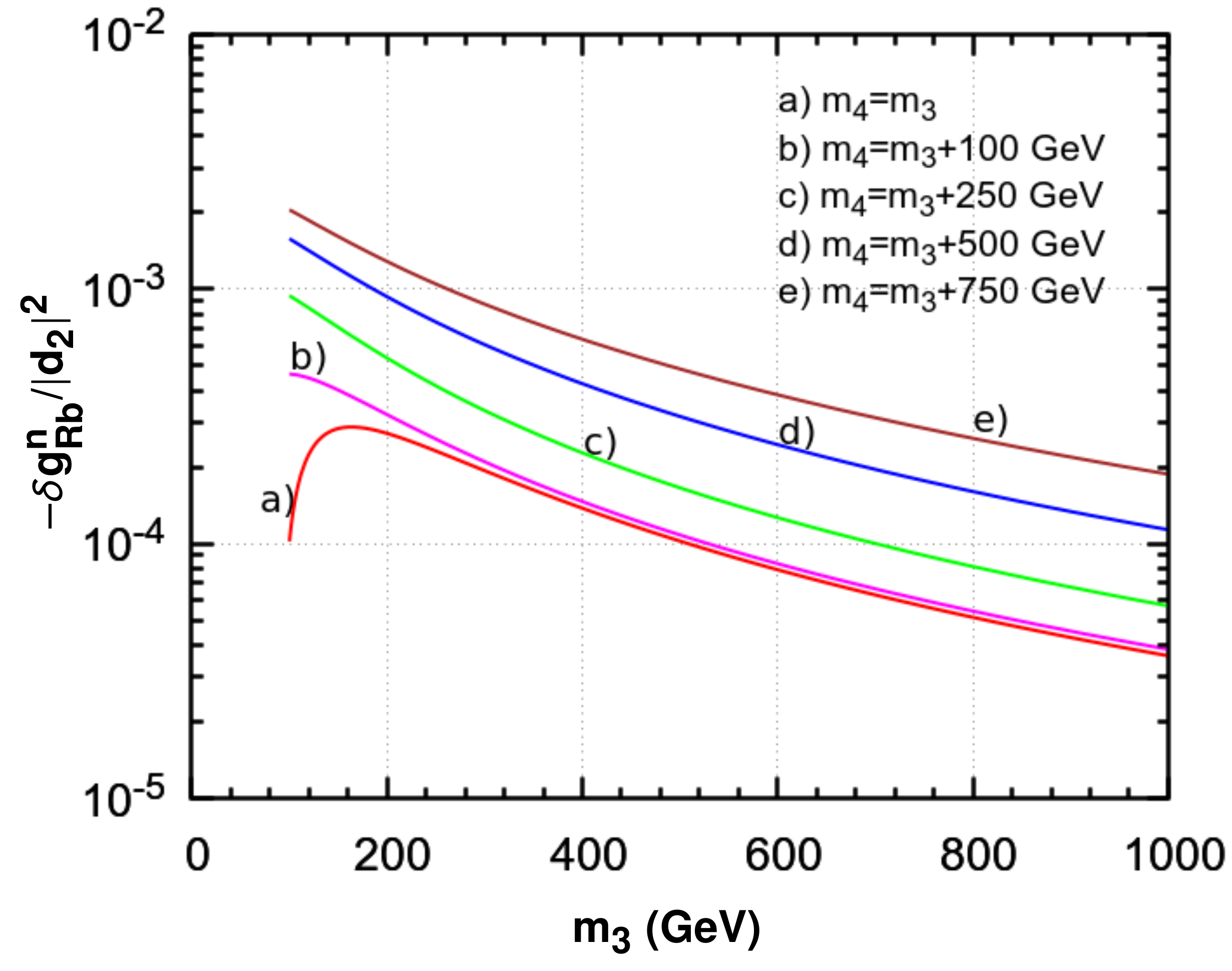}
  \end{tabular}
  \caption{The contributions of the neutral scalars
      to $\delta g^n_{Lb}$ and $\delta g^n_{Rb}$
      as functions of $m_3$,
      for different values of $m_4 - m_3$.}
  \label{hufdodi}
\end{figure*}

We have evaluated the exact Eqs.~\eqref{mcdlls}
by using
\texttt{LoopTools}~\cite{Hahn:1998yk}.\footnote{It is convenient
  to substitute the zeros in many arguments of the Passarino--Veltman
  functions by some small nonzero squared masses,
  lest \texttt{LoopTools} is driven to spurious numerical instabilities.}
We have checked in the numerical simulation that the divergences indeed cancel,
by verifying that the results are independent
of the $\Delta$ parameter of \texttt{LoopTools}. 
Without loss of generality,
we have required that $m_4 > m_3$.
The results are shown in Fig.~\ref{hufdodi}.
It is seen that $\delta g_{Lb}^n > 0$ but $\delta g_{Rb}^n < 0$
(recall that a negative $\delta g_{Rb}$
goes in the wrong direction if one wishes
to explain $A_b$ below the SM value);
both are typically $\mathrm{O} \left( 10^{-4} \right) \left| d_2 \right|^2$
unless $m_3 \sim 200$\,GeV and $m_4 \sim 1$\,TeV,
in which case they may reach
$\mathrm{O} \left( 10^{-3} \right) \left| d_2 \right|^2$.

Comparing Figs.~\ref{fig:2hdm-charged} and~\ref{hufdodi},
one sees that,
unless the masses of the two NP neutral scalars are close to each other,
\emph{there is in general no rationale
for neglecting the neutral-scalar contribution
{as compared to the charged-scalar one.}}

\begin{figure*}[htb]
  \centering
  \begin{tabular}{cc}
    \includegraphics[width=0.47\textwidth]{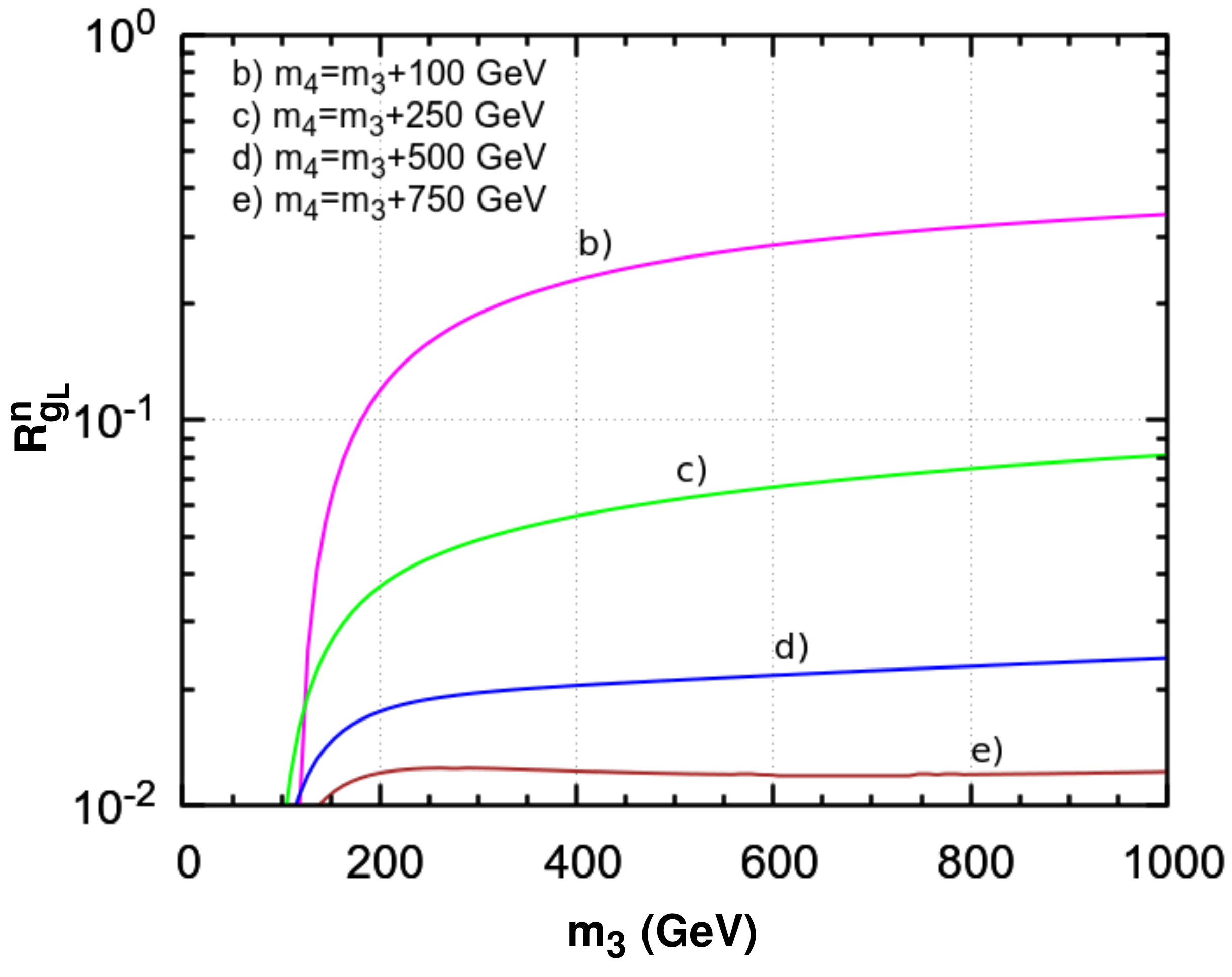}& 
    \includegraphics[width=0.47\textwidth]{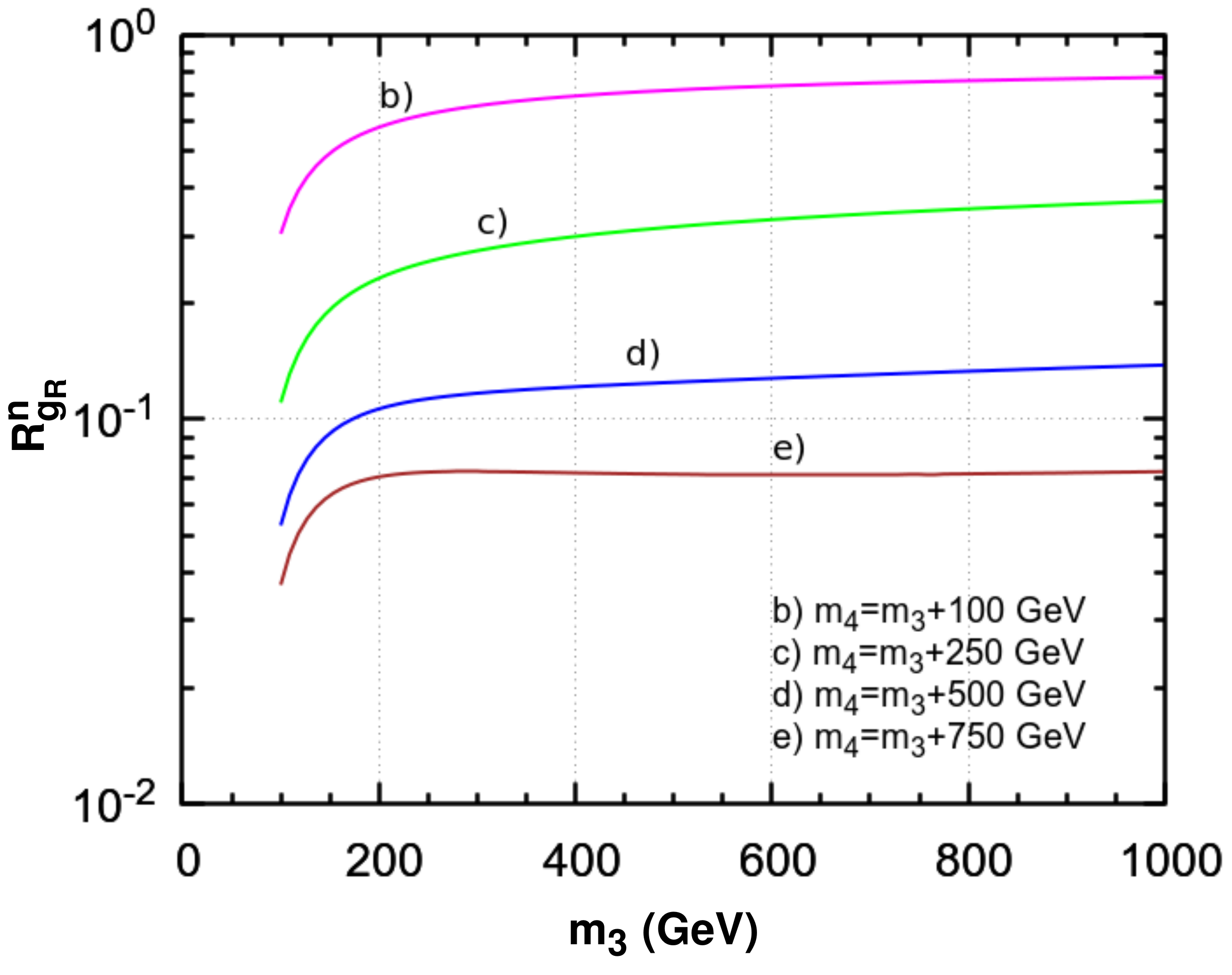}
  \end{tabular}
  \caption{The asymmetries $R_{g_{Lb}}^n$ (left panel)
      and $R_{g_{Rb}}^n$ (right panel),
      defined in a fashion analogous to Eqs.~\eqref{eq:1},
      plotted as functions of $m_3$ for various values of $m_4 - m_3$.}
  \label{Asymmetries-Neutral}
\end{figure*}

We have checked the validity of the approximation
of neglecting $M_Z$ for the case of the neutral scalars.
This is shown in Fig.~\ref{Asymmetries-Neutral}.
We see that
the relative error of neglecting $M_Z$
is much larger in the case of the neutral scalars
than in the case of the charged scalars
(\textit{cf.}\ Fig.~\ref{fig:2hdm-charged-assymetry}),
and it is
larger for $g_R$ than for $g_L$.
{(When} $m_3=m_4$ the asymmetries are 1,
because the approximate expression of Eq.~(\ref{eq:approx})
vanishes for $m_3 = m_4$ while the exact results are nonzero.
We have not
{displayed this case in Fig.~\ref{Asymmetries-Neutral},}
because it would correspond to the upper line in the axes box.)
On the other hand,
the relative error is large precisely when the absolute values
of $\delta g_{Lb}$ and $\delta g_{Rb}$ are small,
\textit{i.e.}\ when the exact values are not very relevant anyway. 

\begin{figure*}[htb]
  \centering
  \begin{tabular}{cc}
    \includegraphics[width=0.45\textwidth]{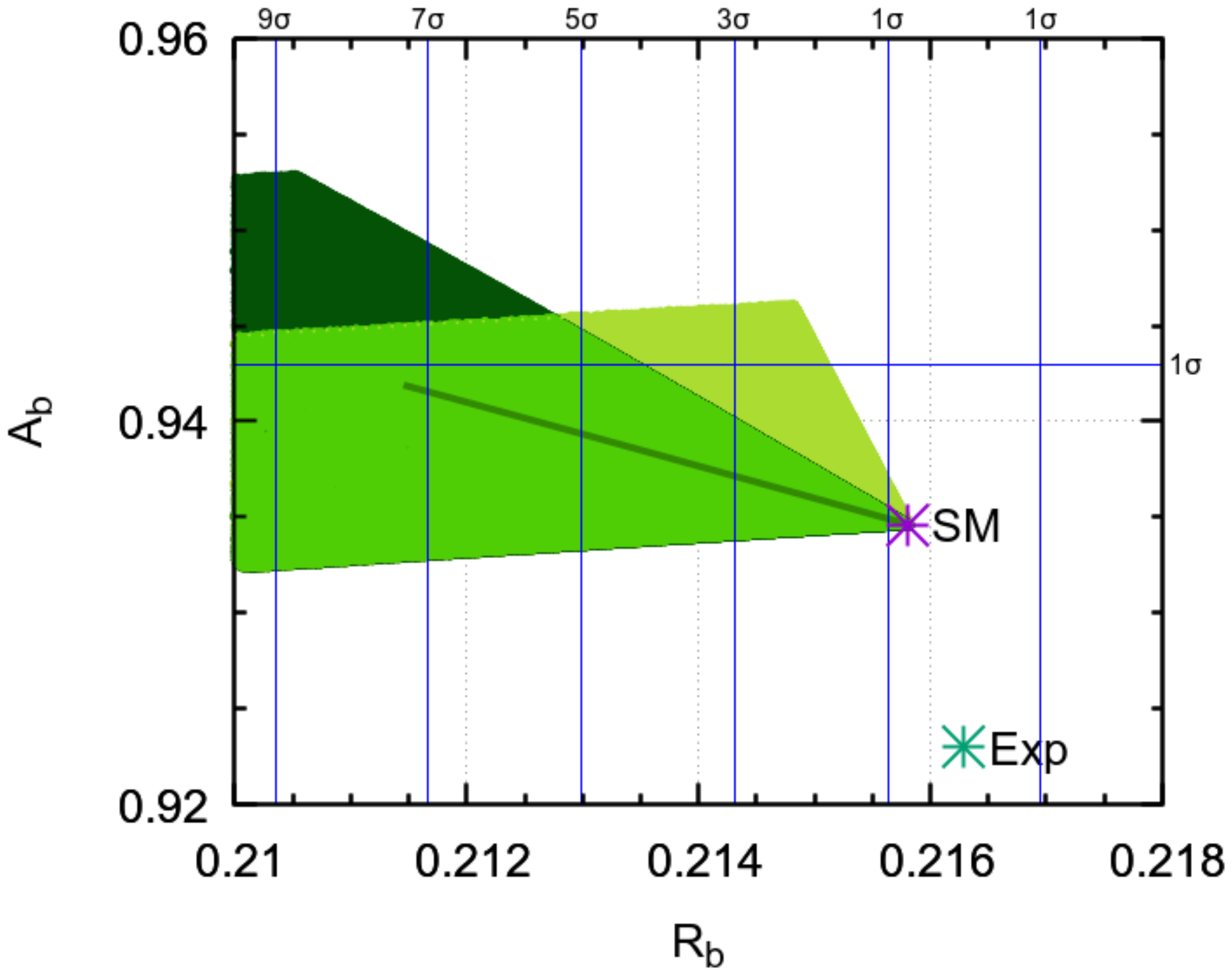} &
    \includegraphics[width=0.45\textwidth]{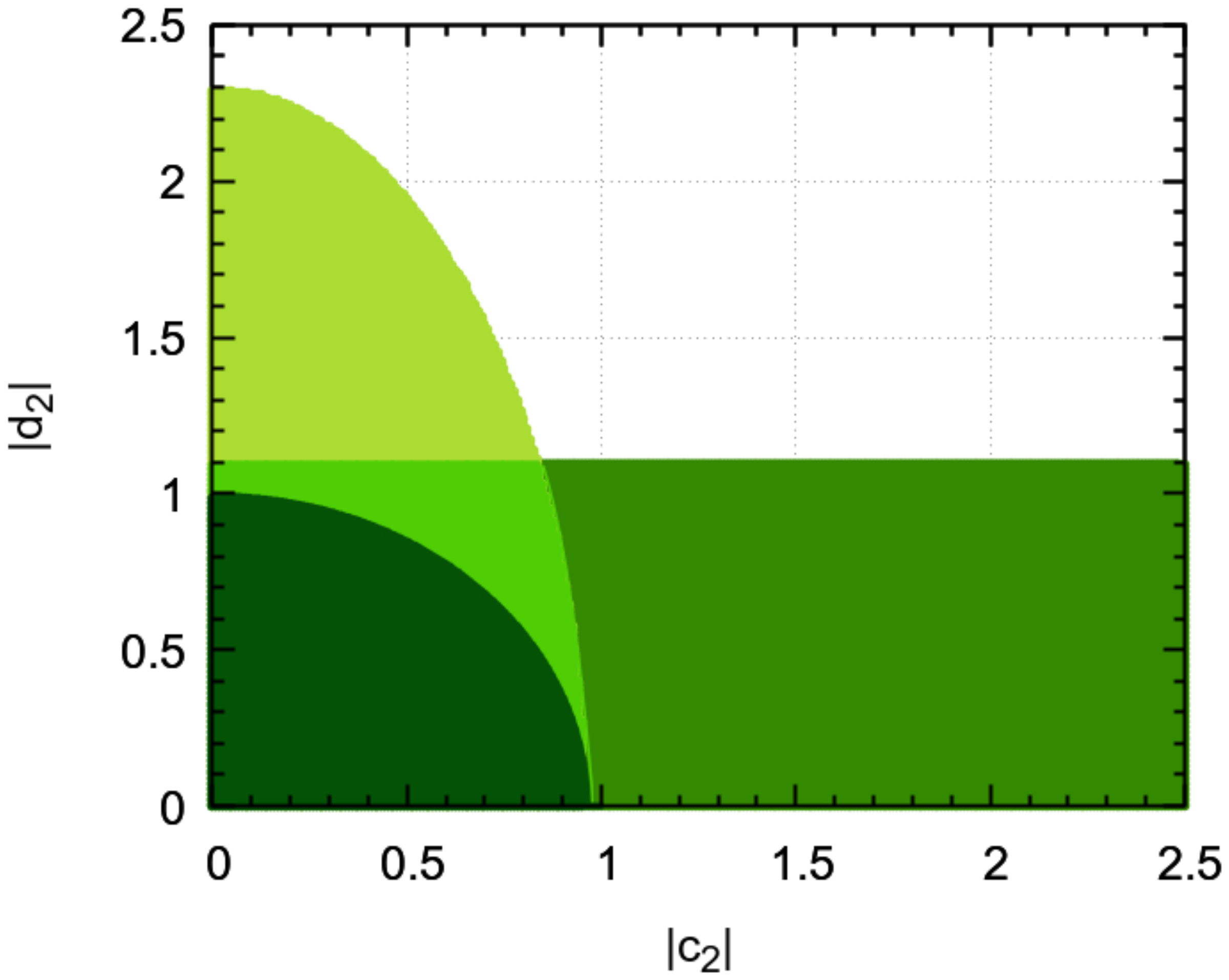}
  \end{tabular}
  \caption{In making the left panel,
      we have used scalar masses $M_{H^+} = 254$\,GeV,
      $m_3 = 250$\,GeV,
      and $m_4 = 850$\,GeV,
      and we have let the Yukawa couplings
      $\left| c_2 \right|$ and $\left| d_2 \right|$ vary in between 0 and 2.5.
      We have depicted the values of $R_b$ and $A_b$
      due to the charged-scalar contribution (in yellowish green),
      the neutral-scalar contribution (a straight line,
      because it is just a function of $\left| d_2 \right|$),
      and the sum of both (in dark and bright green).
      We have also marked the experimental central point (green star),
      the various $n\sigma$ limits (blue lines),
      and the Standard Model prediction (violet star).
      In making the right panel,
      we have used the same scalar masses as for the left panel,
      and we have shown the impact of the $R_b$ $2\sigma$ limits
      on $\left| c_2 \right|$ and $\left| d_2 \right|$;
      the allowed ranges are depicted with
      only the charged-scalar contribution (yellowish and bright green),
      only the neutral-scalar contribution (horizontal band)
      and the sum of both (dark green).
     }
  \label{ChargedAndNeutrals}
\end{figure*}

In the left panel of Fig.~\ref{ChargedAndNeutrals}
we have displayed the impact of both the charged- and neutral-scalar
contributions in the $A_b$--$R_b$ plane.
In making Fig.~\ref{ChargedAndNeutrals}, we have taken into account
the experimental limits, $0.04 < T < 0.20$, on the electroweak
parameter $T$. The 
contribution of the scalars to $T$ is
\be
\label{eq:2}
T = \frac{1}{16\pi s_W^2 M_W^2} \left[
  f \left( M_{H^+}, m_3 \right) + f \left( M_{H^+}, m_4 \right)
  - f \left( m_3, m_4 \right)
  \right],
\ee
where $f \left( x, y \right)$ is the function
\begin{equation}
  \label{eq:3}
  f(x,y)= \frac{x^2 + y^2}{2} - \frac{x^2 y^2}{x^2 - y^2}\,
  \ln\frac{x^2}{y^2}. 
\end{equation}
The function $f$ is zero when $x=y$. In order to keep $T$
sufficiently small, we have set $M_H^+ = 254$ GeV rather close
to $m_3 = 250$ GeV; on the other hand, we
have set $m_4 = 850$\,GeV much larger than
$m_3$, so that $\delta g_{L,Rb}^n$ are rather large, \textit{cf.}\
Fig.~\ref{hufdodi}.
We see that the impact on $A_b$ is always small,
but the impact on $R_b$ may be quite strong when the Yukawa couplings
$c_2$ and $d_2$ become large.  This of course puts bounds on $\left| c_2
\right|$ and $\left| d_2 \right|$, and those bounds are displayed in
the right panel of Fig.~\ref{ChargedAndNeutrals}, using as input the
$2\sigma$ experimental lower bound on $R_b$.  We see that the impact
of the neutral-scalar contributions can be quite drastic,
\textit{cf.}\ the large difference between the dark-green and
light-green areas in the right panel of Fig.~\ref{ChargedAndNeutrals}.

\subsection{The complex 2HDM}

The complex 2HDM (C2HDM) is a two-Higgs-doublet model with a softly
broken $\mathbbm{Z}_2$ symmetry.  The scalar potential is
\ba
V_H &=&
m_{11}^2\, \Phi_1^\dagger \Phi_1
+ m_{22}^2\, \Phi_2^\dagger \Phi_2
- m_{12}^2\, \Phi_1^\dagger \Phi_2
- {m_{12}^2}^\ast\, \Phi_2^\dagger \Phi_1
\no 
& & + \frac{\lambda_1}{2}\, \Phi_1^\dagger \Phi_1\, \Phi_1^\dagger \Phi_1
+ \frac{\lambda_2}{2}\, \Phi_2^\dagger \Phi_2\, \Phi_2^\dagger \Phi_2
+ \lambda_3\, \Phi_1^\dagger \Phi_1\, \Phi_2^\dagger \Phi_2
\no 
& & + \lambda_4\, \Phi_1^\dagger \Phi_2\, \Phi_2^\dagger \Phi_1
+ \! \frac{\lambda_5}{2} \left( \Phi_1^\dagger \Phi_2 \right)^2
+ \! \frac{\lambda_5^\ast}{2} \left( \Phi_2^\dagger \Phi_1 \right)^2 \! \! \!,
\label{VH}
\ea
where all the parameters, except $m_{12}^2$ and $\lambda_5$, are real.
In general, $\textrm{Im} \left[ \left( m_{12}^2 \right)^2
\lambda_5^\ast \right]$ is allowed to be nonzero.  By rephasing
$\Phi_1$ and $\Phi_2$, we go to a basis where the VEVs are real and
positive: $\left\langle 0 \left| \varphi_k^0 \right| 0 \right\rangle =
v_k \left/ \sqrt{2} \right.$ for $k = 1, 2$.  We write
\be
v_1 = v\, c_\beta, \quad \quad v_2 = v\, s_\beta,
\label{real_vevs}
\ee
%
where $v = 246$\,GeV and $0 < \beta < \pi/2$.
Thenceforth, $c_\theta$, $s_\theta$, and
$t_\theta$ represent the cosine, sine, and tangent, respectively, of
whatever angle $\theta$ is in the subindex.  We write the scalar
doublets as
\be
\Phi_k = \left(
\begin{array}{c}
\varphi_k^+\\
\left( v_k + \eta_k + i \chi_k \right) \left/ \sqrt{2} \right.
\end{array}
\right)
\quad \quad (k = 1, 2).
\ee
We transform the fields into the so-called Higgs basis
through~\cite{Botella:1994cs}
\be
\left( \begin{array}{c} H_1 \\ H_2 \end{array} \right)
= \left( \begin{array}{cc}
c_{\beta} & s_{\beta} \\ - s_{\beta} & c_{\beta} \end{array} \right)
\left( \begin{array}{c} \Phi_1 \\ \Phi_2 \end{array} \right).
\ee
Then $H_2$ does not have a VEV:
\bs
\ba
  H_1
&=&
  \left(
    \begin{array}{c} G^+ \\
      \left( v + H^0 + i G^0 \right) \left/ \sqrt{2} \right.
    \end{array}
  \right),
  \\
H_2
&=&
  \left(
    \begin{array}{c}
      H^+ \\
      \left( R_2 + i I_2 \right) \left/ \sqrt{2} \right.
    \end{array}
  \right).
\ea
\es
$G^+$ and $G^0$ are the Goldstone bosons.
There is a charged-scalar pair $H^\pm$ with mass $m_{H^\pm}$.

In a standard C2HDM notation, $\eta_3 := I_2$ and the neutral mass
eigenstates are obtained from the three neutral components as
\be
\left(
\begin{array}{c}
S^0_2\\*[1mm]
S^0_3\\*[1mm]
S^0_4
\end{array}
\right)
= R
\left(
\begin{array}{c}
\eta_1\\
\eta_2\\
\eta_3
\end{array}
\right).
\label{h_as_eta}
\ee
The orthogonal matrix $R$ diagonalizes the neutral mass matrix
\be
\left( {\cal M}^2 \right)_{ij} =
\frac{\partial^2 V_H}{\partial \eta_i\, \partial \eta_j},
\ee
through
\be
R\, {\cal M}^2\, R^T = \textrm{diag} \left( m_2^2, m_3^2, m_4^2 \right),
\ee
where\footnote{In this subsection
we assume that the observed particle with mass 125\,GeV
is \emph{the lightest}\/ neutral scalar.}
$m_2 = 125\, \mathrm{GeV} \leq m_3 \leq m_4$
are the masses of the neutral scalars ($m_1$ is the unphysical mass of
the Goldstone boson $S^0_1=G^0$).
In our numerical study we use $m_{3,4} \in \left[ 125\, \mathrm{GeV},\
800\, \mathrm{GeV} \right]$ with $m_3 < m_4$.  We parameterize the
orthogonal matrix $R$ as~\cite{ElKaffas:2007rq}
\be 
R = \left( \begin{array}{ccccc}
  c_{\alpha_1} c_{\alpha_2} & & s_{\alpha_1} c_{\alpha_2} & & s_{\alpha_2} \\
  - s_{\alpha_1} c_{\alpha_3} - c_{\alpha_1} s_{\alpha_2} s_{\alpha_3} & &
  c_{\alpha_1} c_{\alpha_3} - s_{\alpha_1} s_{\alpha_2} s_{\alpha_3}  & &
  c_{\alpha_2} s_{\alpha_3} \\
  s_{\alpha_1} s_{\alpha_3} - c_{\alpha_1} s_{\alpha_2} c_{\alpha_3} & &
  - c_{\alpha_1} s_{\alpha_3} - s_{\alpha_1} s_{\alpha_2} c_{\alpha_3} & &
  c_{\alpha_2} c_{\alpha_3}
\end{array} \right).
\label{matrixR}
\ee
Without loss of generality,
the angles may be restricted to~\cite{ElKaffas:2007rq}
\be
- \pi/2 < \alpha_1 \leq \pi/2,
\quad 
- \pi/2 < \alpha_2 \leq \pi/2,
\quad
0 \leq \alpha_3 \leq \pi/2.
\label{range_alpha}
\ee
Taking the limit $\alpha_2, \alpha_3 \rightarrow 0$ one recovers a
2HDM with softly broken $\mathbbm{Z}_2$ symmetry and no CP violation;
this is the `real 2HDM', in which $S^0_4=A$ is the massive CP-odd
scalar.

In practice,
because of the experimental limit $1.1 \times 10^{-29}$\,e.cm
on the electric dipole moment of the electron,
both $\alpha_1$ and $\alpha_2$ are much more restricted
than in inequalities~\eqref{range_alpha}:
$\left| \alpha_2 \right| \lesssim 0.1$
and $\alpha_1$ is always very close to $\beta$.

Comparing with Eqs.~\eqref{jdkpsos} and~\eqref{def:U&V}, we find
\bs
\label{eq:700}
\ba
\mathcal{U} &=& \left( \begin{array}{cc}
  c_{\beta} & - s_{\beta} \\ s_{\beta} & c_{\beta}
\end{array} \right),
\\
\mathcal{V} &=& \left( \begin{array}{cccc}
  i c_\beta & R_{11} - i s_\beta R_{13} & R_{21} - i s_\beta R_{23} &
  R_{31} - i s_\beta R_{33}
  \\
  i s_\beta & R_{12} + i c_\beta R_{13} & R_{22} + i c_\beta R_{23} &
  R_{32} + i c_\beta R_{33}
\end{array} \right).
\ea
\es
Equation~\eqref{eq:202} still holds and
\ba
& & \mathrm{Im} \left( \mathcal{V}^\dagger \mathcal{V} \right)
\no &=&
{\footnotesize \left( \begin{array}{cccc}
  0 & - c_\beta R_{11} - s_\beta R_{12} & - c_\beta R_{21} - s_\beta R_{22} &
  - c_\beta R_{31} - s_\beta R_{32} \\
     c_\beta R_{11} + s_\beta R_{12} & 0
    & c_\beta R_{31} + s_\beta R_{32} &
  - c_\beta R_{21} - s_\beta R_{22} \\
    c_\beta R_{21} + s_\beta R_{22}
     & - c_\beta R_{31} - s_\beta R_{32} & 0 &
  c_\beta R_{11} + s_\beta R_{12} \\
  c_\beta R_{31} + s_\beta R_{32} & c_\beta R_{21} + s_\beta R_{22} &
  - c_\beta R_{11} - s_\beta R_{12} & 0
  \end{array} \right).}
\no & &
\ea
Assuming the Yukawa couplings to follow the type-II 2HDM pattern,
\textit{viz.}\ $e_1 = f_2 = 0$ and
\be
e_2 =  \frac{\sqrt{2} m_t}{v_2},
\quad \quad
f_1 =  \frac{\sqrt{2} m_b}{v_1},
\ee
we have
\begin{equation}
\label{eq:207}
c_2 =  \frac{\sqrt{2} m_t}{v} \cot{\beta}, \quad \quad
d_2 = - \frac{\sqrt{2} m_b}{v} \tan{\beta}.
\end{equation}
Note that,
contrary to the assumptions in the previous subsection,
here $\left| c_2 \right|$ and $\left| d_2 \right|$
may be of vastly different orders of magnitude---in particular,
$\left| d_2 \right| \ll \left| c_2 \right|$ for $\tan{\beta} \sim 1$.
However,
when $\tan{\beta} \gtrsim \sqrt{m_t / m_b} \approx 6$,
$\left| d_2 \right|$ becomes larger than $\left| c_2 \right|$,
and that is the regime that we will be mostly interested in.

This model was studied in detail in Ref.~\cite{Fontes:2017zfn}, which
introduced the code {\tt C2HDM\_HDECAY} implementing the C2HDM in {\tt
  HDECAY} \cite{Djouadi:1997yw, Djouadi:2018xqq}.
For illustrative purposes, we take points from that fit, where,
invoking constraints from Flavour Physics on $R_b$\cite{Haber:1999zh},
$\tan{\beta}$ was taken above $0.8$.  In that scan the following
ranges were considered:
  \begin{itemize}
  \item
   $\tan\beta \in [0.8:35]$,
  \item
  $ m_2= 125\text{GeV},\ m_3,m_4 \in [125:800]\text{GeV\,}$,
  \item
  $ M_{H^+} \in [580:800] \text{GeV\,}$,
  \end{itemize}
where $m_4>m_3$ and the constraint on the charged
  Higgs mass comes from B-physics\cite{Deschamps:2009rh, Mahmoudi:2009zx, Misiak:2017bgg,Mahmoudi:2017mtv}
All points passed both the theoretical constraints on
unitarity~\cite{Kanemura:1993hm,Ginzburg:2005dt},
bounded from below,
and the electroweak parameters $S,T,U$,
as well the experimental constraints coming from the LHC.
We combine these with the results from a new dedicated run
$\tan{\beta} \in [0:100]$.
Such extreme (very low and very high) values of
$\tan{\beta}$ may be in 
contradiction with certain Flavour Physics observables, notably (as we
will now show) $Z \rightarrow b \bar{b}$.\footnote{Moreover,
  both extremely high and extremely low values of $\tan\beta$
  will also violate perturbativity.}
Nevertheless, we will consider those extreme values
since we wish to 
stress that the details of such a bound may require
{both the charged-scalar
  and the neutral-scalar contributions.}  As shown in Fig.~8 of
Ref.~\cite{Fontes:2014tga}, very large $\tan{\beta}$ is only
consistent with current measurements at LHC if $\alpha_1$ lies in a
very restricted range $\alpha_1 \approx \beta$, which we impose in
this run \textit{ab initio}.  Moreover, in order to obtain agreement
with the measured EDMs, $\alpha_2$ always turns out to be very small.

As in the alignment
case discussed previously, the contribution
due to the charged Goldstone bosons decouples, it is included in the
SM and subtracted out, and the result from charged scalars is still given by
Eqs.~\eqref{eq:99999} and Fig.~\ref{fig:2hdm-charged}.  Note that
$\delta g_{Lb}^c$ is positive while $\delta g_{Rb}^c$ is negative.  Recall
that the positive $\delta g_{Lb}$
tends to make $R_b$ smaller and from there comes
a bound in the $m_{H^\pm}$--$\tan{\beta}$ plane.
The correction $\delta g_{Rb}^c$ is too small to have an impact on
$R_b$ (see Eq.~\eqref{eq:75a}) but it could have a substantial impact on $A_b$
going in the \emph{wrong} direction when compared with the
experimental measurements (see Eq.~\eqref{eq:75b}).
{However, we will see below
  (see the right panel of Fig.~\ref{fig:delg-compare-high_tb})}
that this only
happens for large values of $\tan\beta$ not allowed by perturbativity.

\begin{figure*}[htb]
  \centering
  \includegraphics[width=0.47\textwidth]{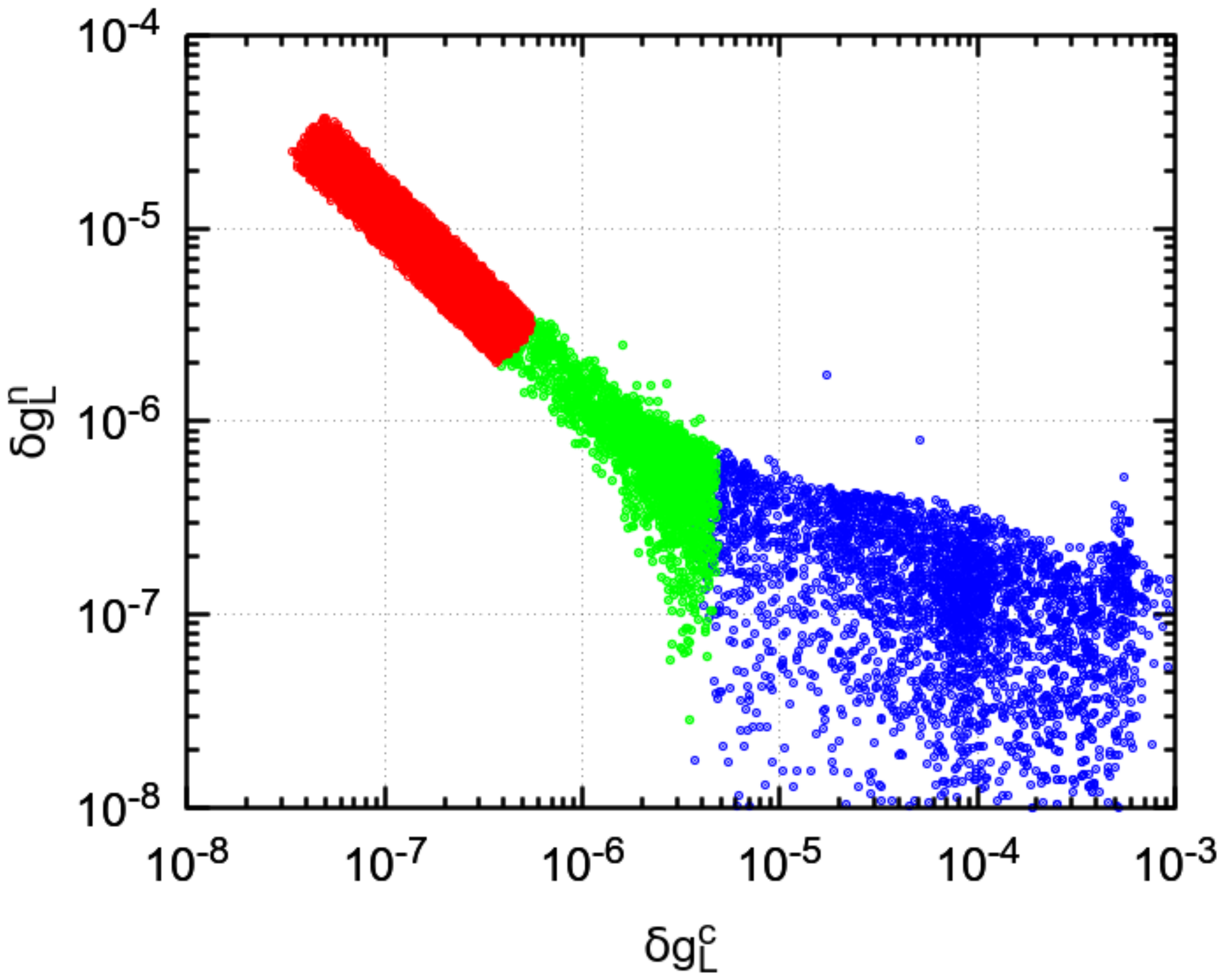}
  \includegraphics[width=0.47\textwidth]{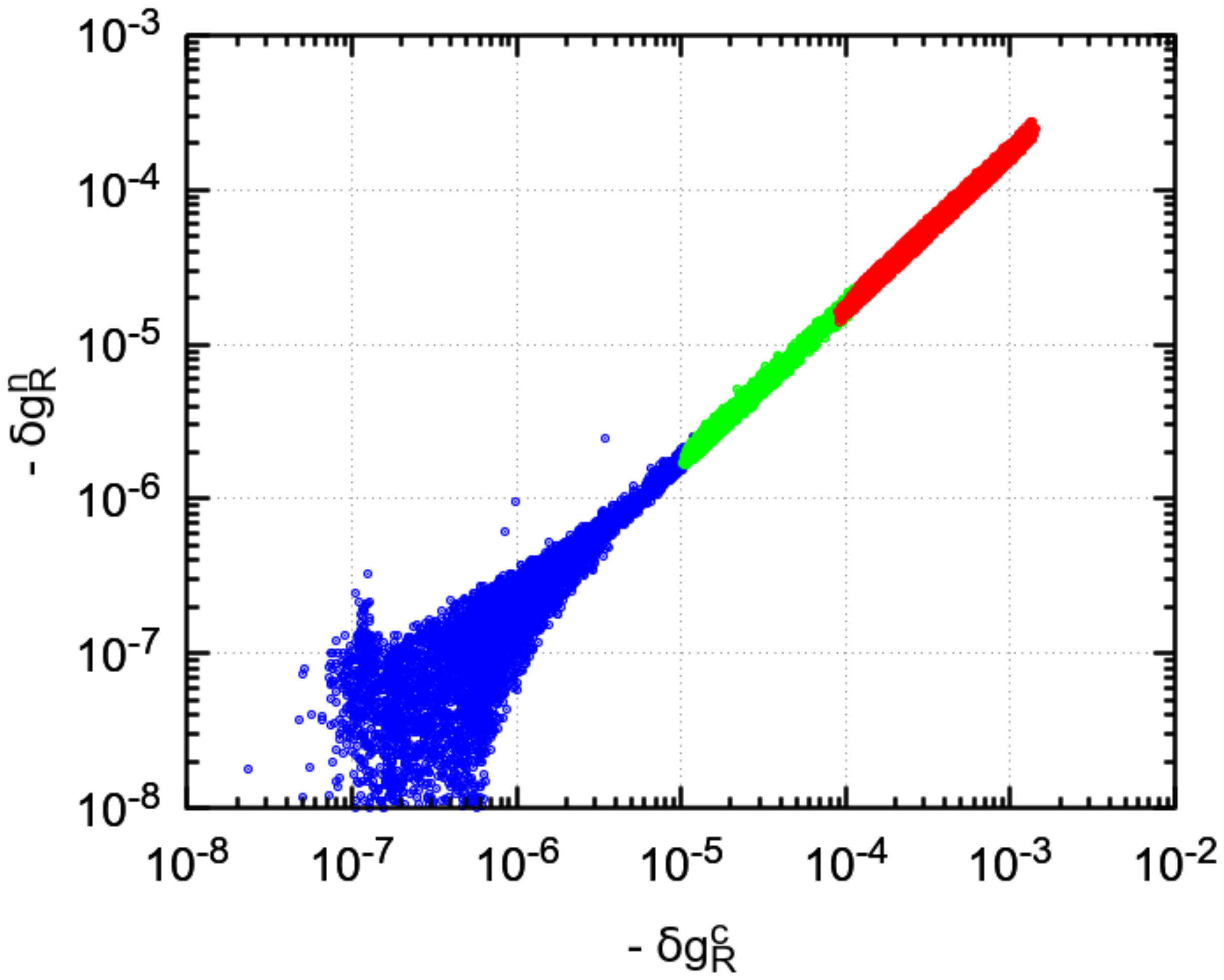}
  \caption{Comparison of $\delta g_{Lb}^n$ with $\delta g_{Lb}^c$ (left plot)
    and of $\delta g_{Rb}^n$ with $\delta g_{Rb}^c$ (right plot).}
  \label{fig:delg-compare-low_tb}
\end{figure*}

We are particularly interested in the contributions to $\delta g_{Lb}$
and $\delta g_{Rb}$ arising from the neutral scalars, because in the
literature they are {frequently disconsidered}.
We would like to know
under which circumstances those contributions can be large.  We have
separated the data of our scans in three different sets:
\begin{itemize}
\item Small $\tan\beta \in [0,10]$, blue in the plots.
\item Intermediate $\tan\beta \in [10,30]$, green in the plots.
\item Large $\tan\beta \in [30,100]$, red in the plots.
\end{itemize}
In the left panel of Fig.~\ref{fig:delg-compare-low_tb}
we display $\delta g_{Lb}^n$ versus $\delta g_{Lb}^c$ for all three sets;
in the right panel,
$-\delta g_{Rb}^n$ is displayed against $-\delta g_{Rb}^c$
(remember that both $\delta g_{Rb}^n$ and $\delta g_{Rb}^c$ are negative).
We see that $\left| \delta g_{Rb}^n \right|$
generally is of order $\left. \left| \delta g_{Rb}^c \right| \right/ 10$,
but they may be comparable in the low-$\tan{\beta}$ regime.
On the other hand,
$\delta g_{Lb}^n \ll \delta g_{Lb}^c$ for low $\tan{\beta}$
but $\delta g_{Lb}^n \gg \delta g_{Lb}^c$ for high $\tan{\beta}$;
they are comparable for $\tan{\beta} \sim 30$.
Thus,
{\emph{one cannot neglect the neutral-scalar contributions
    when $\tan{\beta} \gtrsim 10$.}}
For low $\tb \sim 1$,
$\delta g_{Lb}^c$ is much larger than $\delta g_{Lb}^n$,
but $\delta g_{Rb}^n$ may not be much smaller than $\delta g_{Rb}^c$.

\begin{figure*}[htb]
  \centering
  \includegraphics[width=0.47\textwidth]{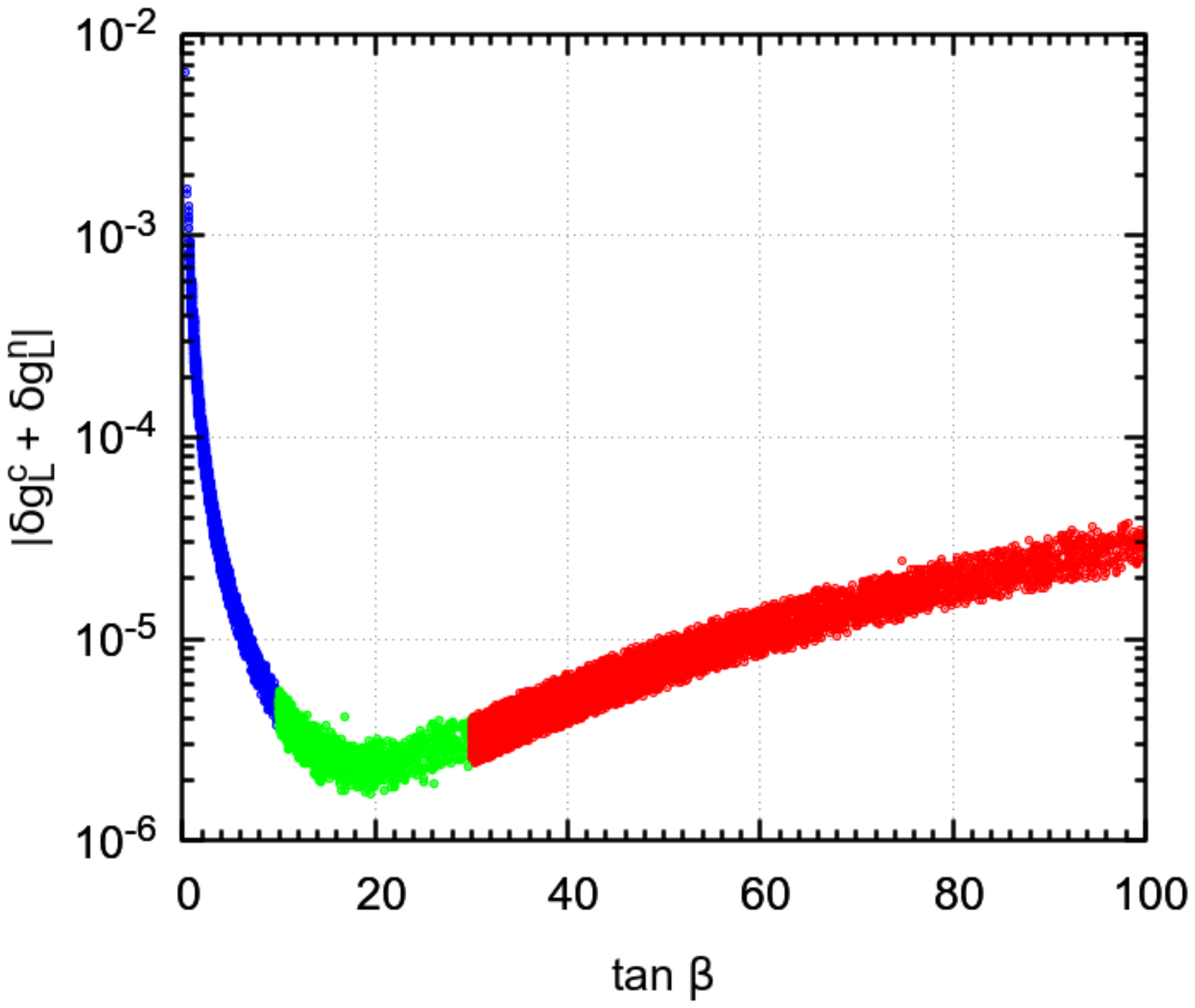}
  \includegraphics[width=0.47\textwidth]{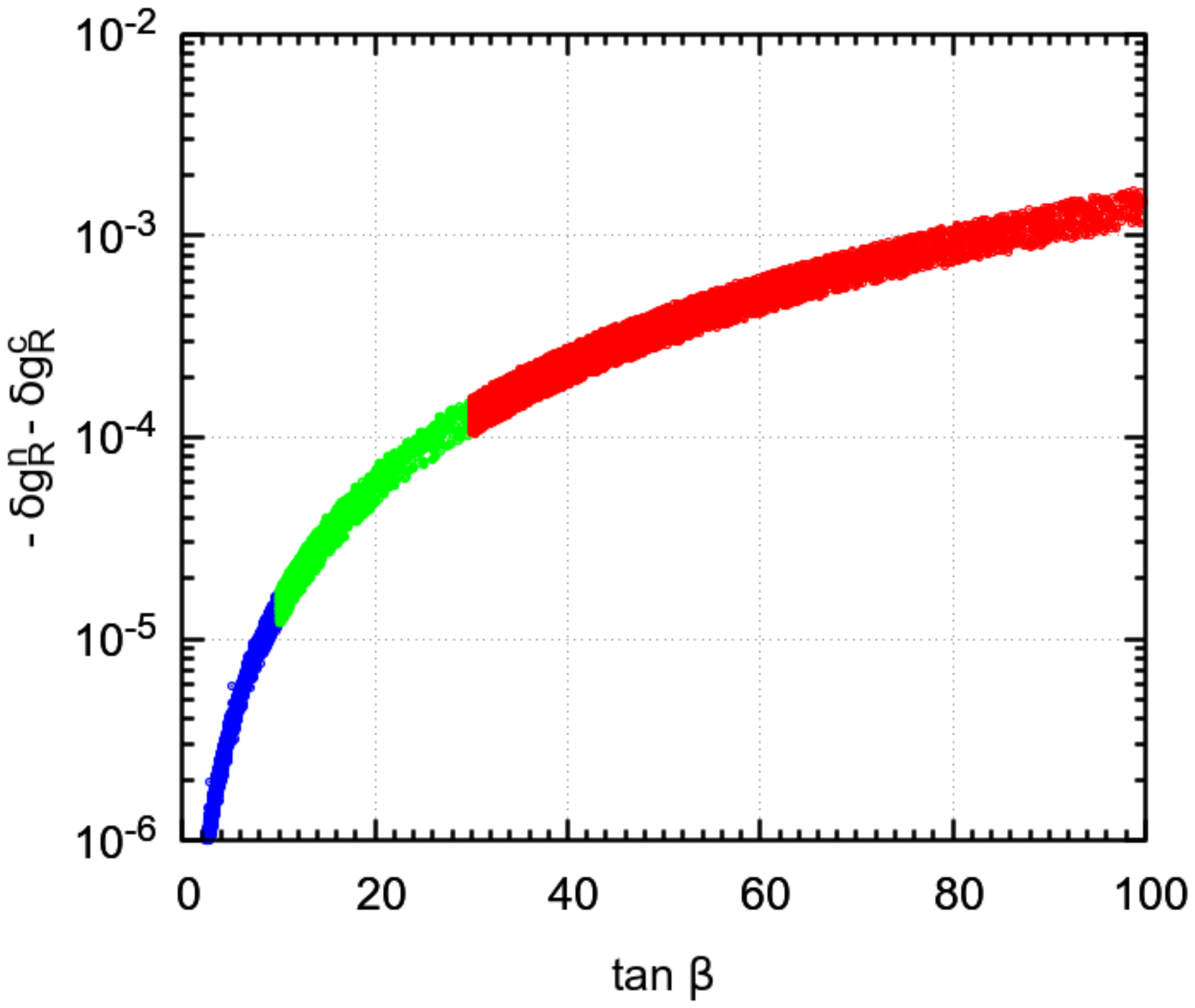}
  \caption{Total contribution of the neutral and charged scalars to
    $\delta g_{Lb}$ and $\delta g_{Rb}$.}
  \label{fig:delg-sums-low_tb}
\end{figure*}

\begin{figure*}[htb]
  \centering
  \includegraphics[width=0.47\textwidth]{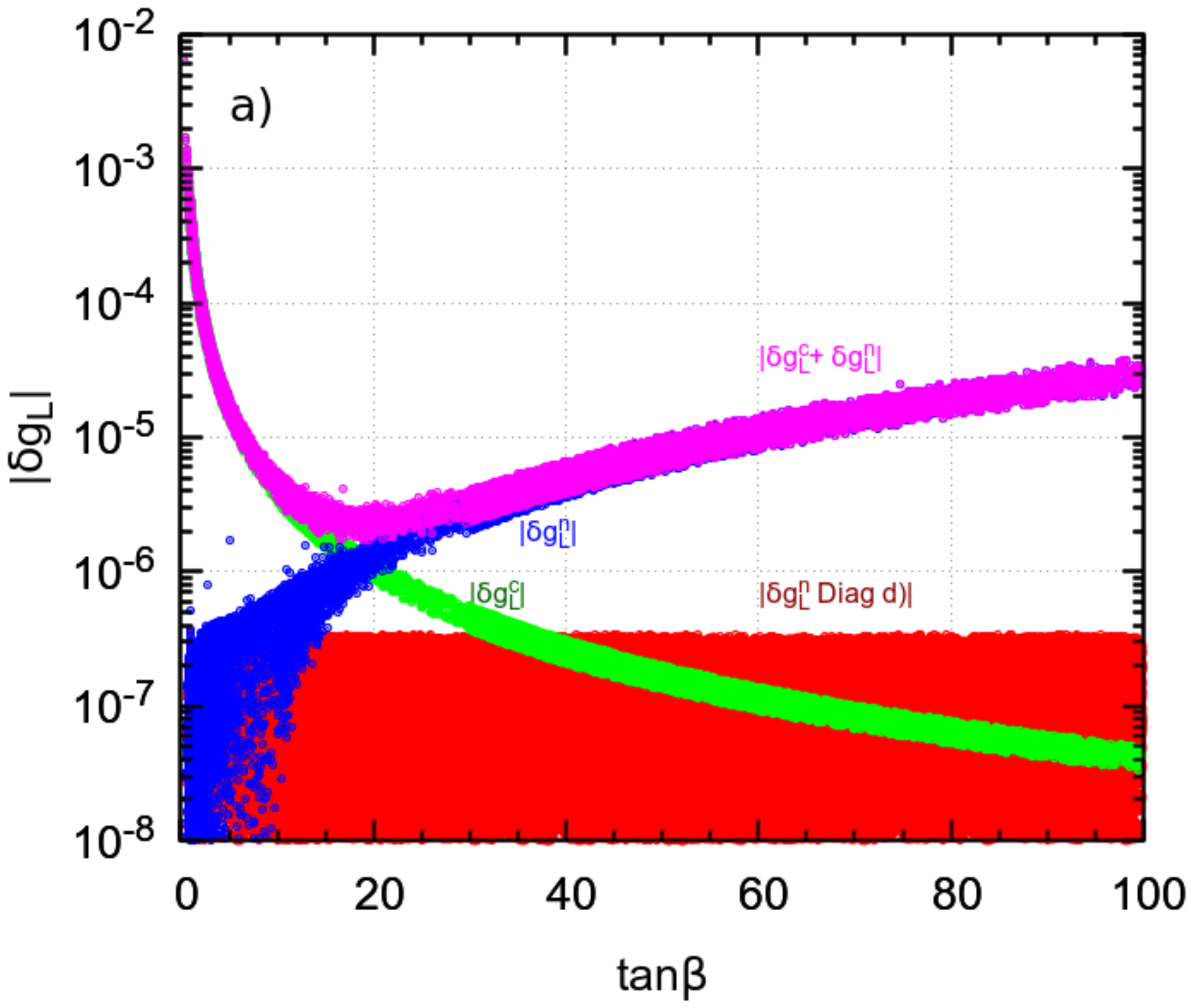}
  \includegraphics[width=0.47\textwidth]{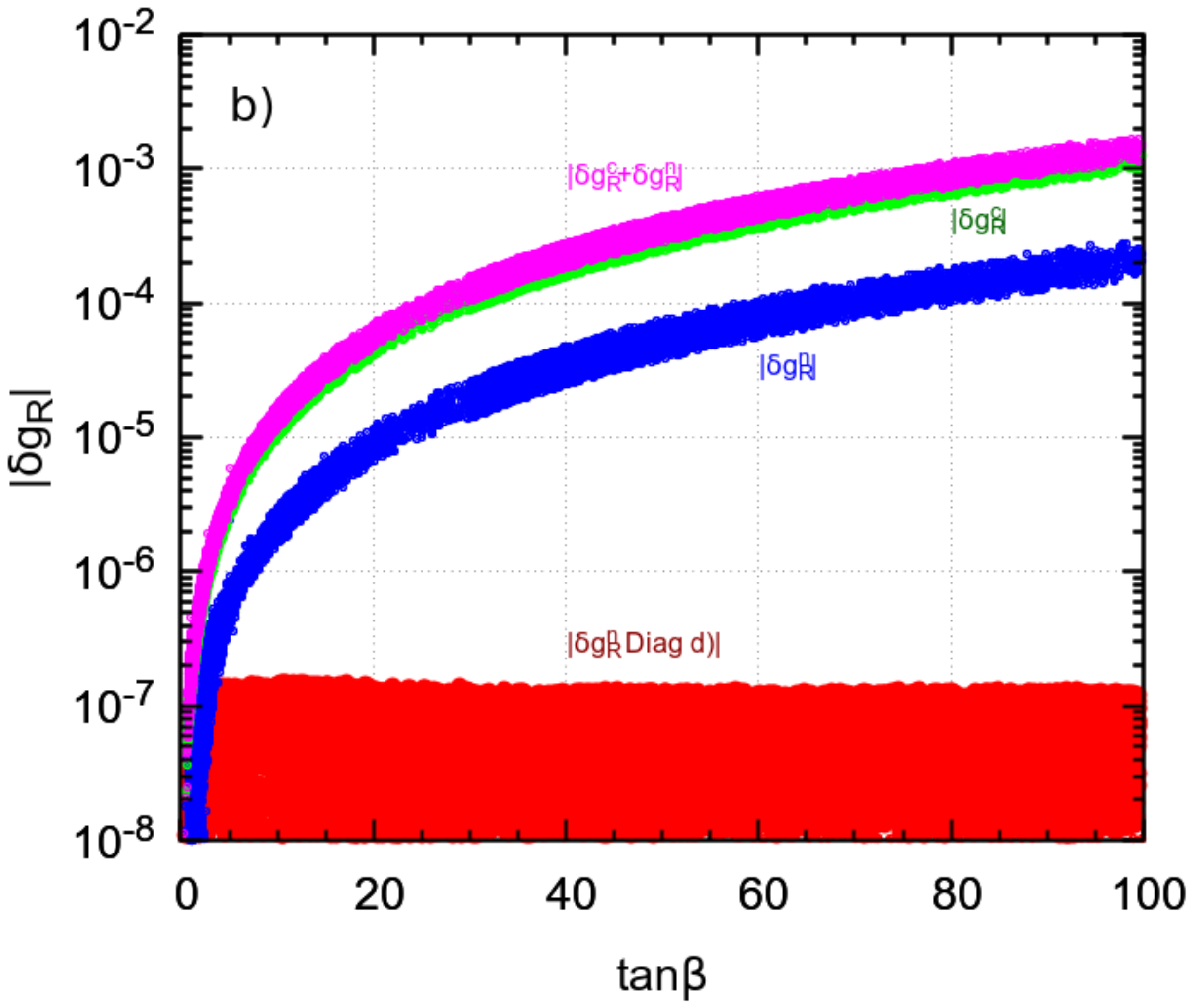}
  \caption{Left panel: comparison of the neutral-scalars contribution
    $\delta g_{Lb}^n$ (in blue)
    and of the charged-scalar contribution $\delta g_{Lb}^c$ (in green)
    with $\delta g_{Lb} = \delta g_{Lb}^n + \delta g_{Lb}^c$ (in pink).
    Right panel: comparison of the neutral-scalars contribution
    $\delta g_{Rb}^n$ (in blue)
    and of the charged-scalar contribution $\delta g_{Rb}^c$ (in green)
    with $\delta g_{Rb} = \delta g_{Rb}^n + \delta g_{Rb}^c$ (in pink).
    Also displayed (in red) are the contributions
    of the diagrams in Fig.~\ref{fig:type_d)}c),d)
    to both $\delta g_{Lb}^n$ (in the left panel)
    and $\delta g_{Rb}^n$ (in the right panel).}
  \label{fig:delg-compare-high_tb}
\end{figure*}

The sums $\delta g_{Lb}^c + \delta g_{Lb}^n$
and $- \delta g_{Rb}^c - \delta g_{Rb}^n$
are displayed as functions of $\tb$ in Fig.~\ref{fig:delg-sums-low_tb}.
We see that a significant impact on $A_b$ and $R_b$
can only occur for either very low or very high values of $\tan\beta$;
namely,
for $\tb \lesssim 1$,
$\delta g_{Lb}^c + \delta g_{Lb}^n \sim 10^{-3}$,
and for $\tb \gtrsim 50$,
$- \delta g_{Rb}^c - \delta g_{Rb}^n \gtrsim 10^{-3}$.

Both Figs.~\ref{fig:delg-compare-low_tb} and~\ref{fig:delg-sums-low_tb}
are depicted together in Fig.~\ref{fig:delg-compare-high_tb}.
In particular,
in Fig.~\ref{fig:delg-compare-high_tb}a) we see that
the neutral-scalar contribution to $\delta g_{Lb}$
becomes larger than the charged-scalar contribution,
eventually by many orders of magnitude,
as soon as $\tan{\beta} > 30$.
Thus,
\emph{one cannot neglect the contribution of the neutral scalars
  to $\delta g_{Lb}$}.
We expect this effect to be even more important
in models with more than two Higgs doublets and/or extra singlets.

It is interesting to inquire about the importance of the type d)
neutral-scalar contributions (red in Fig.~\ref{fig:delg-compare-high_tb}).
One sees that,
when $\tb$ is low,
they may constitute a substantial part of the $\delta g_{\aleph b}^n$
($\aleph = L, R$),
but that is precisely the range when the $\delta g_{\aleph b}^n$
are anyway much too small to be of practical relevance.
We conclude that,
{at least in this particular case},
it is correct to neglect the diagrams in Fig.~\ref{fig:type_d)}c),d),
{as was done in Ref.~\cite{Haber:1999zh}.}

\begin{figure}[!htb]
  \centering
  \includegraphics[width=0.65\textwidth]{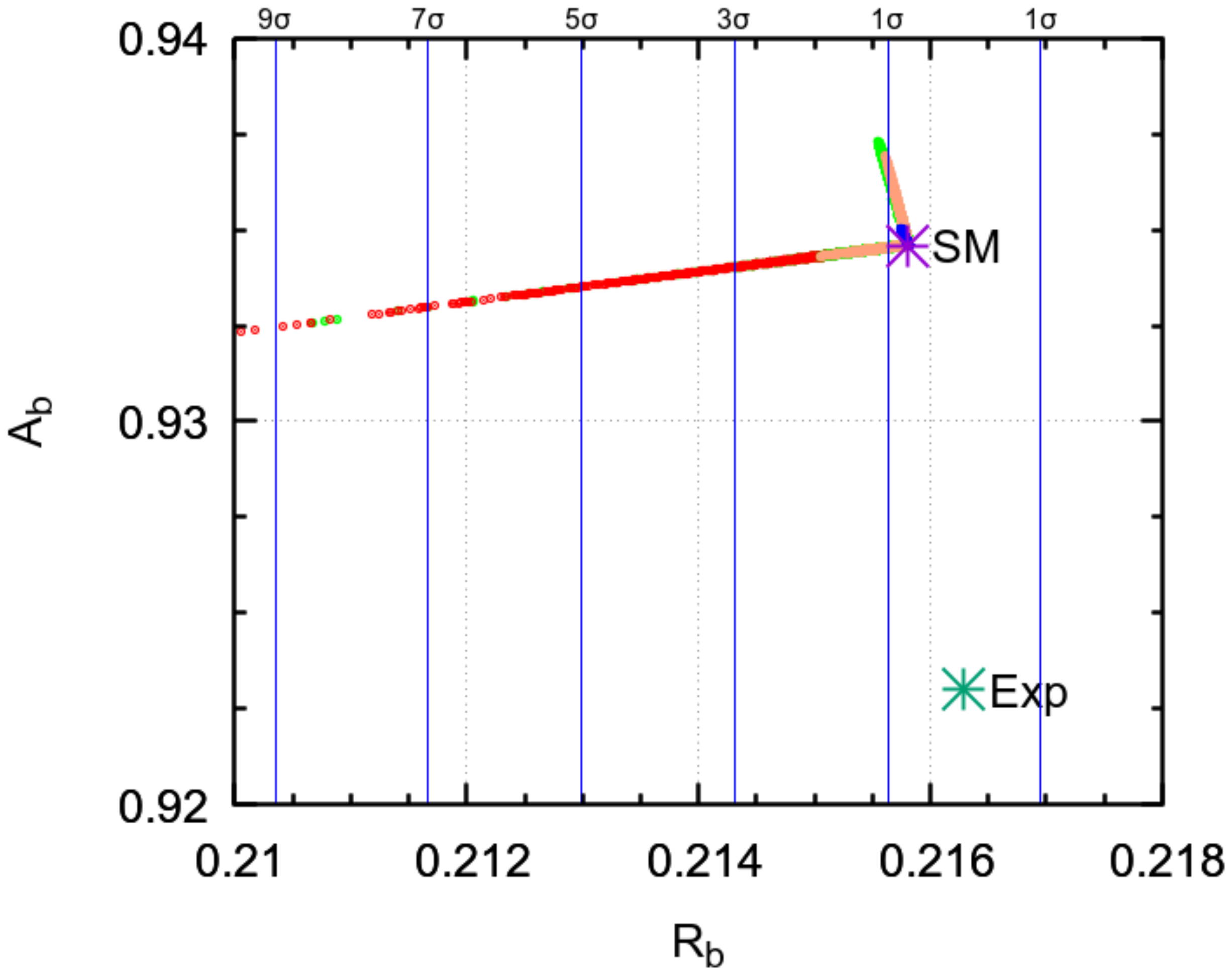}
  \caption{$A_b$ versus $R_b$ in the C2HDM for all values of
    $0<\tan\beta <100$.
    The charged-scalar contribution is shown in
    red for low $\tb$ and in orange for large $\tb$.
    The contribution of the neutral-scalars is in blue and lies very
    close to the SM point.
    In green (in background) the sum of the contributions.}
  \label{fig:RbAb-low_tb-2}
\end{figure}

The impact on $A_b$ and $R_b$ is shown in Fig.~\ref{fig:RbAb-low_tb-2}
for all values of $\tan\beta$ and including the various contributions.
In the low $\tan\beta$ regime,
the charged-scalar contribution (shown in red) is dominant.
The points in Fig.~\ref{fig:RbAb-low_tb-2} only stray
from the 2$\sigma$ $R_b$ bounds for $\tb < 0.8$.
This is the reason why only points with $\tb >0.8$
were taken in Ref.~\cite{Fontes:2017zfn}. In orange is shown the
contribution of the charged scalars for $\tb $ up to 100. The
contribution of the neutral scalars is in blue, and is always very small.
We have verified that for the neutral scalars to
have meaningful impact, one would 
have to consider values of $\tan\beta>250$, which would violate
perturbativity of the Yukawa couplings.\footnote{Although
  in the C2HDM the enhancement of the
  neutral contributions is related to a ratio of
  vevs ($v_2/v_1 = \tan{\beta}$) which is
  limited by perturbativity, in more general models
  where such vev enhancements are less constrained,
  the neutral contributions will be important.
  This can be simulated in the C2HDM by taking
  $\tan{\beta}$ to forbiddingly high values.}
  
We conclude that,
when studying the impact on $Z \rightarrow b \bar{b}$
of multi-scalar models with very large couplings (which means very
large $\tb$ in our example of the C2HDM),
the neutral scalar contributions should be taken into account.
Of course,
{in studying any model one needs to include}
all the theoretical and experimental constraints,
and this may curtail a large part of the phase space for
such extreme couplings.
This will have to be evaluated in a case by case basis.

\section{\label{sec:conclusions}Conclusions}

We have studied the one-loop contributions to $Z \rightarrow b \bar b$
in models with extra scalars.
We have started by deriving the conditions on generic couplings
that must hold for the divergences to cancel.
We have then concentrated on models
with any number of
extra $SU(2)_L$ doublets and singlets,
either neutral,
as in Ref.~\cite{Haber:1999zh},
or charged.
The final expressions are greatly simplified (and very compact),
due to the parameterization
in Refs.~\cite{Grimus:2002ux, Grimus:2007if,Grimus:2008nb, Grimus:1989pu}.
We also extend the analysis in Ref.~\cite{Haber:1999zh}
to models with CP violation in the scalar sector.
We have shown that, in these general models, the conditions previously
derived necessary for the cancellation of the divergences naturally hold. 
  We have then highlighted the possible importance
  of the neutral-scalar contributions.
  In particular, in Fig.~\ref{hufdodi}
  and Fig.~\ref{fig:delg-compare-high_tb}a) we show that,
  in a specific models,
  the contributions of the neutral scalars to $\delta g_{Lb}$
  may in some cases be much larger
  than the contributions of the charged scalars, and this has to be
  considered in evaluating the limits on $A_b$ and $R_b$ as shown, for
  instance, in 
  Fig.~\ref{ChargedAndNeutrals}.

\vspace{1ex}

\section*{Acknowledgments}
We are grateful to A.~Barroso and P.M.~Ferreira for discussions,
and to H.E.~Haber for clarifications on Ref.~\cite{Haber:1999zh}.
This work was supported by the Portuguese
\textit{Funda\c c\~ao para a Ci\^encia e a Tecnologia}\/ (FCT)
through the projects UID/FIS/00777/2019,
PTDC/FIS-PAR/29436/2017,
and CERN/FIS-PAR/
0004/2017;
these projects are partially funded through POCTI (FEDER),
COMPETE,
QREN,
and the EU.
D.F.\ is also supported by FCT under the project SFRH/BD/135698/2018.

\appendix
\section{\label{app:PV}Passarino--Veltman functions}

In this appendix we expose our definition of the Passarino--Veltman
functions, which coincides with that of
\texttt{FeynCalc}~\cite{Mertig:1990an, Shtabovenko:2016sxi} and 
\texttt{LoopTools}~\cite{Hahn:1998yk, vanOldenborgh:1990yc} used in
the algebraic and numerical calculations~\cite{Fontes:2019wqh}. 
We use dimensional regularization;
the Feynman integrals
are performed in a space--time of dimension $d = 4 - \epsilon$.
Then,
\begin{equation}
\label{b1}
\mu^\epsilon \! \int \! \frac{\mathrm{d}^d k}{\left( 2 \pi \right)^d}\
\frac{1}{k^2 - m_0^2}\
\frac{1}{\left( k + r \right)^2 - m_1^2}\ k^\lambda
= \frac{i}{16 \pi^2}\ r^\lambda\, B_1 \left( r^2, m_0^2, m_1^2 \right).
\end{equation}
Moreover,
\ba
& & \mu^\epsilon \! \int \! \frac{\mathrm{d}^d k}{\left( 2 \pi \right)^d}\
\frac{1}{k^2 - m_0^2}\
\frac{1}{\left( k + r_1 \right)^2 - m_1^2}\
\frac{1}{\left( k + r_2 \right)^2 - m_2^2}
\no 
&=&
\frac{i}{16 \pi^2}\
C_0 \left[ r_1^2, \left( r_1 - r_2 \right)^2, r_2^2, m_0^2, m_1^2, m_2^2 \right].
\label{c0}
\ea
Also,
\ba
& & \mu^\epsilon \! \int \! \frac{\mathrm{d}^d k}{\left( 2 \pi \right)^d}\
\frac{1}{k^2 - m_0^2}\
\frac{1}{\left( k + r_1 \right)^2 - m_1^2}\
\frac{1}{\left( k + r_2 \right)^2 - m_2^2}\ k^\lambda
\no 
&=& \frac{i}{16 \pi^2} \left( r_1^\lambda\, C_1 + r_2^\lambda\, C_2 \right)
\left[ r_1^2, \left( r_1 - r_2 \right)^2, r_2^2,
m_0^2, m_1^2, m_2^2 \right].
\label{c1}
\ea
Finally,
\ba
& & \mu^\epsilon \! \int \! \frac{\mathrm{d}^d k}{\left( 2 \pi \right)^d}\
\frac{1}{k^2 - m_0^2}\
\frac{1}{\left( k + r_1 \right)^2 - m_1^2}\
\frac{1}{\left( k + r_2 \right)^2 - m_2^2}\ \! k^\lambda k^\nu
\no &=& \frac{i}{16 \pi^2}
\left[ g^{\lambda \nu} C_{00} + r_1^\lambda r_1^\nu C_{11} + r_2^\lambda r_2^\nu C_{22}
\right. \no & & \left.
+ \left( r_1^\lambda r_2^\nu + r_2^\lambda r_1^\nu \right) C_{12} \right]
\left[ r_1^2, \left( r_1 - r_2 \right)^2, r_2^2, m_0^2, m_1^2, m_2^2 \right].
\label{c00}
\ea

\vspace{2ex}


\begin{thebibliography}{10}

\bibitem{Aad:2012tfa}
ATLAS, G.~Aad {\em et~al.},
\newblock Phys. Lett. B {\bf 716}, 1 (2012), [1207.7214].

\bibitem{Chatrchyan:2012ufa}
CMS, S.~Chatrchyan {\em et~al.},
\newblock Phys. Lett. B {\bf 716}, 30 (2012), [1207.7235].

\bibitem{gunion:1989we}
J.~F. Gunion, H.~E. Haber, G.~L. Kane and S.~Dawson,
\newblock {\em The Higgs hunter's guide} (Westview Press, 1990),
\newblock Frontiers in Physics.

\bibitem{Branco:2011iw}
G.~C. Branco {\em et~al.},
\newblock Phys. Rept. {\bf 516}, 1 (2012), [1106.0034].

\bibitem{Ivanov:2017dad}
I.~P. Ivanov,
\newblock Prog. Part. Nucl. Phys. {\bf 95}, 160 (2017), [1702.03776].

\bibitem{Romao:2012pq}
J.~C. Romao and J.~P. Silva,
\newblock Int. J. Mod. Phys. {\bf A27}, 1230025 (2012), [1209.6213].

\bibitem{Haber:1999zh}
H.~E. Haber and H.~E. Logan,
\newblock Phys. Rev. {\bf D62}, 015011 (2000), [hep-ph/9909335].

\bibitem{Tanabashi:2018oca}
Particle Data Group, M.~Tanabashi {\em et~al.},
\newblock Phys. Rev. {\bf D98}, 030001 (2018).

\bibitem{Dorsch:2013wja}
G.~Dorsch, S.~Huber and J.~No,
\newblock JHEP {\bf 10}, 029 (2013), [1305.6610].

\bibitem{Basler:2016obg}
P.~Basler, M.~Krause, M.~M\"{u}hlleitner, J.~Wittbrodt, and A.~Wlotzka,
\newblock JHEP {\bf 02}, 121 (2017), [1612.04086].

\bibitem{Krause:2016xku}
M.~Krause, M.~M\"{u}hlleitner, R.~Santos, and H.~Ziesche,
\newblock Phys. Rev. D {\bf 95}, 075019 (2017), [1609.04185].

\bibitem{Fontes:2015gxa}
D.~Fontes, J.~C. Rom\~ao, J.~P. Silva, and R.~Santos,
\newblock {Large pseudo-scalar components in the C2HDM},
\newblock in {\em {2nd Toyama International Workshop on Higgs as a Probe of New
  Physics (HPNP2015) Toyama, Japan, February 11-15, 2015}}, 2015, [1506.00860].

\bibitem{Mader:2012pm}
W.~Mader, J.-H. Park, G.~M. Pruna, D.~St\"ockinger, and A.~Straessner,
\newblock JHEP {\bf 09}, 125 (2012), [1205.2692],
\newblock [Erratum: JHEP 01, 006 (2014)].

\bibitem{Belusca-Maito:2016dqe}
H.~Bélusca-Maïto, A.~Falkowski, D.~Fontes, J.~C. Romão and J.~P. Silva,
\newblock Eur. Phys. J. C {\bf 77}, 176 (2017), [1611.01112].

\bibitem{Campbell:2016zbp}
R.~Campbell, S.~Godfrey, H.~E. Logan and A.~Poulin,
\newblock Phys. Rev. D {\bf 95}, 016005 (2017), [1610.08097].

\bibitem{Hartling:2014aga}
K.~Hartling, K.~Kumar and H.~E. Logan,
\newblock Phys. Rev. D {\bf 91}, 015013 (2015), [1410.5538].

\bibitem{Hartling:2014zca}
K.~Hartling, K.~Kumar and H.~E. Logan,
\newblock Phys. Rev. D {\bf 90}, 015007 (2014), [1404.2640].

\bibitem{Chiang:2014bia}
C.-W. Chiang, S.~Kanemura and K.~Yagyu,
\newblock Phys. Rev. D {\bf 90}, 115025 (2014), [1407.5053].

\bibitem{Degrande:2015xnm}
C.~Degrande, K.~Hartling, H.~E. Logan, A.~D. Peterson and M.~Zaro,
\newblock Phys. Rev. D {\bf 93}, 035004 (2016), [1512.01243].

\bibitem{Tang:2017rhv}
Y.-L. Tang,
\newblock Phys. Rev. D {\bf 97}, 035020 (2018), [1709.07735].

\bibitem{vonBuddenbrock:2018xar}
S.~von Buddenbrock {\em et~al.},
\newblock J. Phys. G {\bf 46}, 115001 (2019), [1809.06344].

\bibitem{Han:2017etg}
L.~Wang, R.~Shi and X.-F. Han,
\newblock Phys. Rev. D {\bf 96}, 115025 (2017), [1708.06882].

\bibitem{Flacher:2008zq}
H.~Fl\"acher {\it et~al.},
\newblock Eur. Phys. J. C {\bf 60}, 543 (2009), [0811.0009],
\newblock [Erratum: Eur.Phys.J.C 71, 1718 (2011)].

\bibitem{Haller:2018nnx}
J.~Haller {\em et~al.},
\newblock Eur. Phys. J. C {\bf 78}, 675 (2018), [1803.01853].

\bibitem{Grimus:2002ux}
W.~Grimus and L.~Lavoura,
\newblock Phys. Rev. D {\bf 66}, 014016 (2002), [hep-ph/0204070].

\bibitem{Grimus:2007if}
W.~Grimus, L.~Lavoura, O.~M. Ogreid and P.~Osland,
\newblock J. Phys. {\bf G35}, 075001 (2008), [0711.4022].

\bibitem{Grimus:2008nb}
W.~Grimus, L.~Lavoura, O.~Ogreid and P.~Osland,
\newblock Nucl. Phys. B {\bf 801}, 81 (2008), [0802.4353].

\bibitem{Grimus:1989pu}
W.~Grimus and H.~Neufeld,
\newblock Nucl. Phys. B {\bf 325}, 18 (1989).

\bibitem{Hollik:1988ii}
W.~F.~L. Hollik,
\newblock Fortsch. Phys. {\bf 38}, 165 (1990).

\bibitem{Hollik:1993cg}
W.~Hollik,
\newblock Adv. Ser. Direct. High Energy Phys. {\bf 14}, 37 (1995).

\bibitem{Christensen:2008py}
N.~D. Christensen and C.~Duhr,
\newblock Comput.Phys.Commun. {\bf 180}, 1614 (2009), [0806.4194].

\bibitem{Nogueira:1991ex}
P.~Nogueira,
\newblock J. Comput. Phys. {\bf 105}, 279 (1993).

\bibitem{Mertig:1990an}
R.~Mertig, M.~B\"ohm, and A.~Denner,  
\newblock Comput. Phys. Commun. {\bf 64}, 345 (1991),
\newblock Available at https://www.feyncalc.org/.

\bibitem{Shtabovenko:2016sxi}
V.~Shtabovenko, R.~Mertig and F.~Orellana,
\newblock Comput. Phys. Commun. {\bf 207}, 432 (2016), [1601.01167].

\bibitem{Fontes:2019wqh}
D.~Fontes and J.~C. Romão,
\newblock Comput. Phys. Commun.  (2020), [1909.05876].

\bibitem{Passarino:1978jh}
G.~Passarino and M.~J.~G. Veltman,
\newblock Nucl. Phys. {\bf B160}, 151 (1979).

\bibitem{Hahn:1998yk}
T.~Hahn and M.~Perez-Victoria,
\newblock Comput. Phys. Commun. {\bf 118}, 153 (1999), [hep-ph/9807565].

\bibitem{vanOldenborgh:1990yc}
G.~van Oldenborgh,
\newblock Comput. Phys. Commun. {\bf 66}, 1 (1991).

\bibitem{Kundu:1995qb}
A.~Kundu and B.~Mukhopadhyaya,
\newblock Int. J. Mod. Phys. {\bf A11}, 5221 (1996), [hep-ph/9507305].

\bibitem{Field:1997gz}
J.~Field,
\newblock Mod. Phys. Lett. A {\bf 13}, 1937 (1998), [hep-ph/9801355].

\bibitem{Gunion:2002zf}
J.~F. Gunion and H.~E. Haber,
\newblock Phys. Rev. D {\bf 67}, 075019 (2003), [hep-ph/0207010].

\bibitem{Botella:1994cs}
F.~Botella and J.~P. Silva,
\newblock Phys. Rev. D {\bf 51}, 3870 (1995), [hep-ph/9411288].

\bibitem{ElKaffas:2007rq}
A.~W. El~Kaffas, P.~Osland and O.~M. Ogreid,
\newblock Nonlin. Phenom. Complex Syst. {\bf 10}, 347 (2007), [hep-ph/0702097].

\bibitem{Fontes:2017zfn}
D.~Fontes {\em et~al.},
\newblock JHEP {\bf 02}, 073 (2018), [1711.09419].

\bibitem{Djouadi:1997yw}
A.~Djouadi, J.~Kalinowski and M.~Spira,
\newblock Comput. Phys. Commun. {\bf 108}, 56 (1998), [hep-ph/9704448].

\bibitem{Djouadi:2018xqq}
A.~Djouadi, J.~Kalinowski, M.~Muehlleitner and M.~Spira,
\newblock Comput. Phys. Commun. {\bf 238}, 214 (2019), [1801.09506].

\bibitem{Deschamps:2009rh}
O.~Deschamps {\em et~al.},
\newblock Phys. Rev. D {\bf 82}, 073012 (2010), [0907.5135].

\bibitem{Mahmoudi:2009zx}
F.~Mahmoudi and O.~Stal,
\newblock Phys. Rev. D {\bf 81}, 035016 (2010), [0907.1791].

\bibitem{Misiak:2017bgg}
M.~Misiak and M.~Steinhauser,
\newblock Eur. Phys. J. C {\bf 77}, 201 (2017), [1702.04571].

\bibitem{Mahmoudi:2017mtv}
F.~Mahmoudi,
\newblock PoS {\bf CHARGED2016}, 012 (2017).

\bibitem{Kanemura:1993hm}
S.~Kanemura, T.~Kubota and E.~Takasugi,
\newblock Phys. Lett. B {\bf 313}, 155 (1993), [hep-ph/9303263].

\bibitem{Ginzburg:2005dt}
I.~Ginzburg and I.~Ivanov,
\newblock Phys. Rev. D {\bf 72}, 115010 (2005), [hep-ph/0508020].

\bibitem{Fontes:2014tga}
D.~Fontes, J.~C. Romão and J.~P. Silva,
\newblock Phys. Rev. {\bf D90}, 015021 (2014), [1406.6080].

\end{thebibliography}

\end{document}